\documentclass[a4paper,11pt]{article}
\pdfoutput=1 

\usepackage{jheppub} 

\usepackage{fancyhdr}
\usepackage{graphicx}
\usepackage{epstopdf}
\usepackage{psfrag}
\usepackage{xspace}
\usepackage{rotating}
\usepackage{cancel}
\usepackage{braket}
\usepackage{slashed}
\usepackage{multicol}

\DeclareSymbolFont{extraup}{U}{zavm}{m}{n}
\DeclareMathSymbol{\varheart}{\mathalpha}{extraup}{86}
\DeclareMathSymbol{\vardiamond}{\mathalpha}{extraup}{87}

\usepackage{amsfonts}
\usepackage{pifont}
\usepackage{multirow}

\newcommand{\SARAH}{{\tt SARAH}\xspace}

\newcommand{\Yes}{\checkmark}

\newcommand{\be}{\begin{equation}}
\newcommand{\ee}{\end{equation}}
\newcommand{\bea}{\begin{eqnarray}}

\newcommand{\eea}{\end{eqnarray}}

\newcommand{\Rmnum}[1]{\expandafter\@slowromancap\romannumeral #1@}

\def\gev{\,{\rm GeV}}
\def\tev{\,{\rm TeV}}

\begin{document}


 \title{\boldmath Features of electroweak symmetry breaking in five dimensional SUSY models}

\author{Zygmunt  Lalak,}
\author{Marek  Lewicki,}
\author{Moritz McGarrie}
\author{and Pawe\l{} Olszewski}
\affiliation{
Institute of Theoretical Physics, Faculty of Physics, University of Warsaw ul. Pasteura 5, 02-093 Warsaw, Poland}
\emailAdd{Zygmunt.Lalak@fuw.edu.pl}
\emailAdd{marek.lewicki@fuw.edu.pl}
\emailAdd{moritz.mcgarrie@fuw.edu.pl}
\emailAdd{pawel.olszewski@fuw.edu.pl}


\abstract{We explore the phenomenological predictions of a supersymmetric standard model, with a large extra dimension and unifying gauge couplings.  The modified five dimensional renormalisation group equations make it possible to obtain light, maximally mixed stops, with a low scale of supersymmetry breaking and a low unification scale. This allows the fine-tuning to be lowered right down to the barrier coming directly from experimental lower limits on the stop masses.
We also show that modifying the SUSY breaking pattern to obtain lighter stops at the high scale does not result in fine-tuning relaxation, and only RGE effects turn out to be effective in generating a lower fine-tuning.}


\maketitle
\flushbottom


\section{Introduction} \label{sec:intro}

The discovery of the Higgs boson, of mass $m_h \sim 125$ GeV, at the first LHC run  \cite{Aad:2012tfa,Chatrchyan:2012ufa} and various null results in searches for super-particles (sparticles) appear to imply that the sparticles of minimal theories of supersymmetry (such as the MSSM) are likely to be out of reach of the LHC altogether. Further considerations regarding electroweak symmetry breaking and the naturalness aesthetic bring further doubt that perhaps supersymmetry is not a symmetry of nature after all.  

At this time it is also worth to consider non minimal models that share much of the well grounded theoretical elegance of the (four dimensional) MSSM: supersymmetry, gauge coupling unification, anomaly cancellation, a minimal matter content to achieve these, unification of matter representations, an explanation of the big hierarchy problem, various dark matter candidates. The model we consider is five dimensional and brings with it additional interesting features: the possibility to observe Kaluza-Klein states of gauge (and other) fields \cite{Antoniadis:1990ew}, the possibility to achieve the observed Higgs mass with sparticles within reach of the next LHC run, and a much lower unification scale and supersymmetry breaking scale than is normally possible in four dimensions.

In this paper we explore electroweak symmetry breaking in a particular class of five dimensional supersymmetric theories \cite{Pomarol:1998sd,Abdalgabar:2015ora}. We also wish to address the possibility to construct a `natural theory' where stops are lighter than their first and second generation counterparts, and which is not spoiled by the renormalisation group effects that would usually make the 3rd generation similarly heavy at a low scale, after a long period of RG-running.
Studying models that differ in how matter fields are located among branes and bulk opens the possibility to explore the effect different SUSY breaking patterns and modified five dimensional RGEs have on fine-tuning wrt a four dimensional theory.
Unfortunately modifying the breaking pattern to obtain more natural soft terms at the high scale does not give the expected fine-tuning relaxation. However we show that modified RGEs enable us to obtain light maximally mixed stops. This allows us to lower fine-tuning down to the barrier coming directly from  observational lower limits on the stop mass. 
 
  In our analysis we used renormalisation group equations outlined in \cite{Abdalgabar:2014bfa,Abdalgabar:2015ora} and adapted a C++ based spectrum generator originally intended for the (four dimensional) MSSM \cite{Lalak:2013bqa}.  A similar modification may be carried out with any publicly available spectrum generator \cite{Allanach:2001kg,Porod:2003um,Djouadi:2002ze} \footnote{To date, five dimensional theories are one such class of models that cannot yet be explored using \SARAH \cite{Staub:2009bi,Staub:2010jh,Staub:2012pb,Staub:2013tta} although it can still be a powerful tool to determine the RGEs of the low energy four dimensional effective theory that the five dimensional theory runs to \cite{Abdalgabar:2015ora}.}. The RGEs used in this paper may be found in  \cite{Abdalgabar:2015ora} and further conventions in \cite{Abdalgabar:2014bfa} and \cite{Hebecker:2001ke,Mirabelli:1997aj,McGarrie:2010kh,Cornell:2011fw,McGarrie:2013hca}. 
For earlier phenomenological studies of five dimensional theories see for example \cite{Bhattacharyya:2010rm}.

\par The outline of the paper is as follows:  section \ref{sec:model} we outline the two base models that we wish to explore.  In section \ref{sec:running} we explore the behaviour of the running parameters of the theories. In section \ref{sec:ewsb} we look at naturalness and electroweak symmetry breaking as we vary the radius of the extra dimension.  We also look at benchmark models of supersymmetry breaking in section \ref{sec:benchsusy}. In section \ref{sec:natural} we outline and explore a ``natural'' model in which the 3rd generation are spatially located on a different brane to the first two generations and in which supersymmetry is gauge mediated directly to the first two generations but only indirectly to the 3rd,  which ideally would allow for light stops.  In section \ref{sec:conclude} we conclude.  In appendix \ref{sec:Numerical}  we outline the implementation of spectrum generator and RG solver in C++ code.

\section{The 5D-SSM+$( F^{\pm})$ Model}\label{sec:model}

\begin{figure}[!th]
\begin{center}
\includegraphics[width=4cm,angle=0]{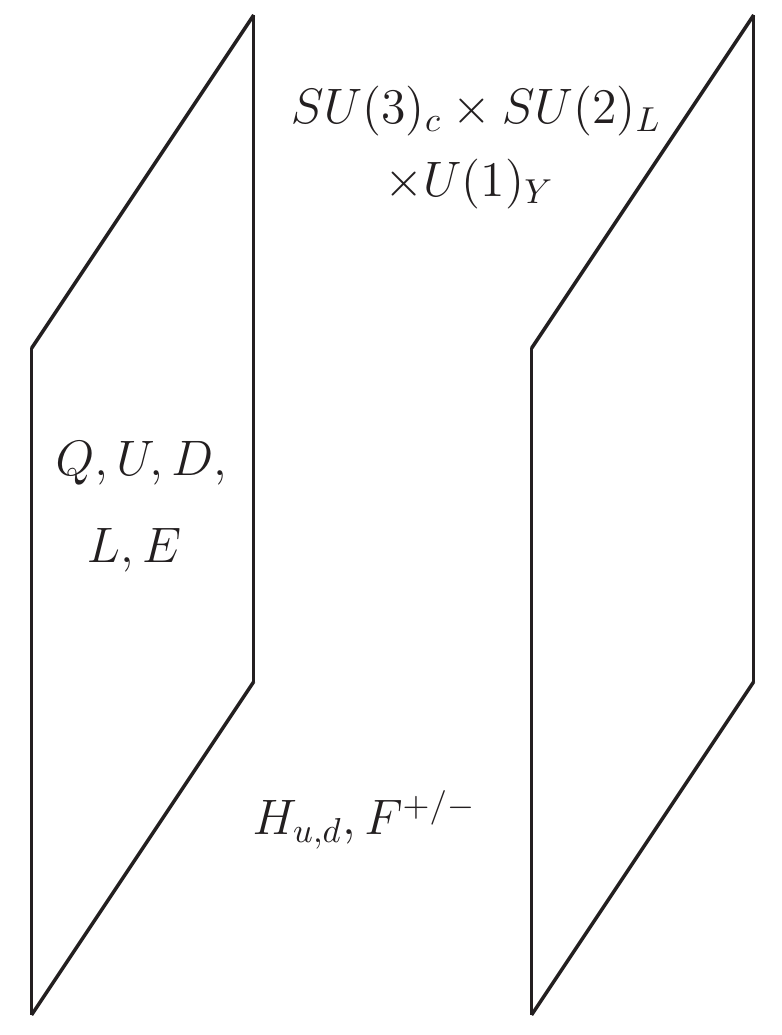}\hspace{3cm}
\includegraphics[width=4cm,angle=0]{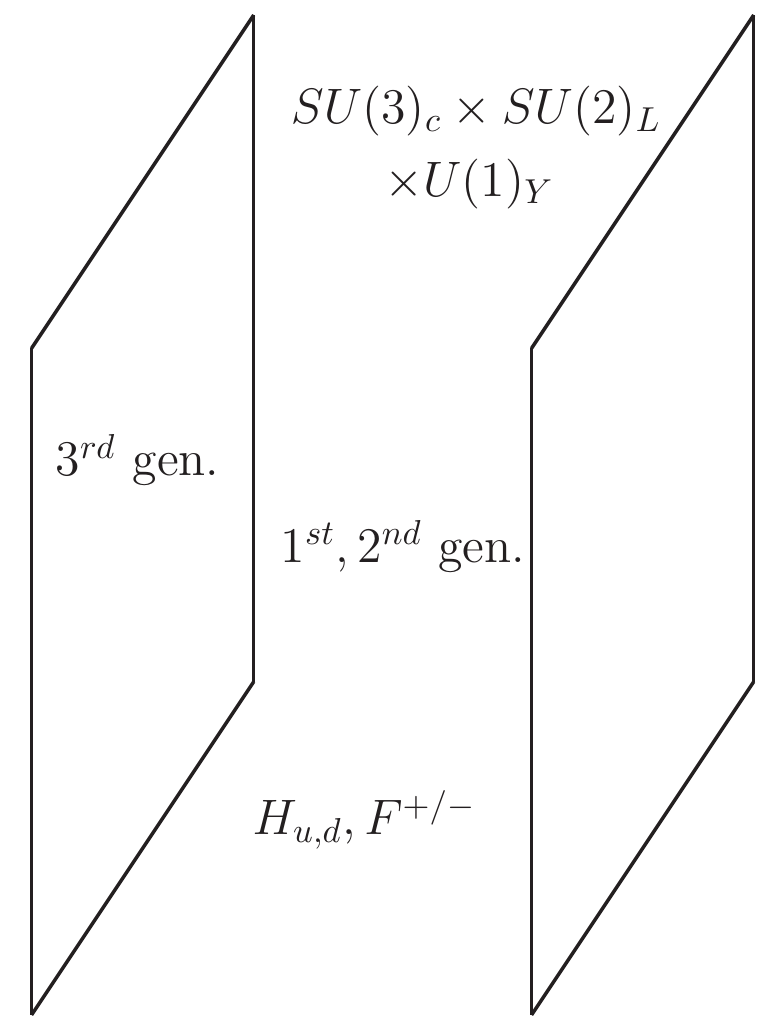}
\caption{{\it 
Pictorials to represent the location of matter in the five dimensional model.  In Model 1  (left), all generations of matter live on a brane. In Model 2 (right), only the 3rd generation live on a brane
}}  
\label{fig:models}
\end{center}
\end{figure}

The first model that we wish to explore is a five dimensional supersymmetric theory with the field content outlined in table \ref{Table:Model1} and is pictured in figure \ref{fig:models} (left).  In this model the Higgs fields ($H_u,H_d$), gauge fields and additionally $F^{\pm}$ are bulk fields \cite{Delgado:1998qr}. This matter content is necessary  for the gauge couplings unification, as we shall explore further later.
All five dimensional bulk matter fields are supersymmetric Hypermultiplets which due to even and odd boundary conditions lead to a four dimensional Chiral multiplet as a zero mode of the Kaluza-Klein expansion: such details are well documented, for instance in \cite{Hebecker:2001ke,Mirabelli:1997aj,McGarrie:2010kh,Cornell:2011fw,McGarrie:2013hca}.   The second model we wish to explore is outlined in table \ref{Table:Model2} and pictured in figure \ref{fig:models}  (right). In model 2 only the third generation is located on a brane and the first and second generation are in the bulk along with the Higgs multiplets and $F^{\pm}$ fields.

\begin{table}
\begin{center} 
\begin{tabular}{|c|c|c|c|c|} 
\hline 
Superfields & Brane& Bulk & $U(1)_Y\times SU(2)_L \times SU(3)_c$ \\
\hline 
\(\hat{q}^{f}\) &\Yes&-&  \((\frac{1}{6},{\bf 2},  {\bf 3}) \)  \\ 
\(\hat{d}^{f}\) &\Yes&-& \((\frac{1}{3},
{\bf 1},  {\bf \overline{3}}) \)  \\ 
\(\hat{u}^{f}\) &\Yes&-& \((-\frac{2}{3},
{\bf 1},  {\bf \overline{3}}) \)  \\
\(\hat{l}^{f}\) & \Yes&-& \((-\frac{1}{2}, {\bf 2}, {\bf 1}) \) \\
\(\hat{e}^{f}\) & \Yes&-& \((1, {\bf 1},
 {\bf 1}) \)  \\  \hline\hline
\(\hat{H}_d\)   & - &\Yes  & \( (-\frac{1}{2},{\bf 2},  {\bf 1}) \) \\ 
\(\hat{H}_u\) &  - &\Yes      & \( (\frac{1}{2},{\bf 2},  {\bf 1}) \)  \\ 
 \hline
\(\hat{F}_-\)   & - &\Yes     & \( (-1,{\bf 1},  {\bf 1}) \) \\ 
\(\hat{F}_+\) & - &\Yes       & \( (1,{\bf 1},  {\bf 1}) \)  \\ 
 \hline\hline
\(\hat{B}_V\)   & - &\Yes     & \( (0,{\bf 1},  {\bf 1}) \) \\ 
\(\hat{W}_V\) & - &\Yes       & \( (0,{\bf 3},  {\bf 1}) \)  \\ 
\(\hat{G}_V\)   & - &\Yes     & \( (0,{\bf 1},  {\bf 8}) \) \\ 
 \hline\hline
\end{tabular} \caption{The matter content of model 1.  All superfields of chiral fermions live on a brane and all Higgs-type superfields and gauge vector fields live in the bulk.  The superscript $f=1,2,3$ denotes the generations. Neutrino superfields may be included straightforwardly. 
\label{Table:Model1}}
\end{center} 
\end{table}

\begin{table}
\begin{center} 
\begin{tabular}{|c|c|c|c|c|} 
\hline 
Superfields & Brane& Bulk & $U(1)_Y\times SU(2)_L \times SU(3)_c$ \\
\hline 
\(\hat{q}^{1,2}\) & - &\Yes&  \((\frac{1}{6},{\bf 2},  {\bf 3}) \)  \\ 
\(\hat{d}^{1,2}\) & - &\Yes& \((\frac{1}{3},
{\bf 1},  {\bf \overline{3}}) \)  \\ 
\(\hat{u}^{1,2}\)& - &\Yes& \((-\frac{2}{3},
{\bf 1},  {\bf \overline{3}}) \)  \\
\(\hat{l}^{1,2}\) & - &\Yes& \((-\frac{1}{2}, {\bf 2}, {\bf 1}) \) \\
\(\hat{e}^{1,2}\)& - &\Yes& \((1, {\bf 1},
 {\bf 1}) \)  \\ 
\hline 
\(\hat{q}^{3}\) &\Yes &-&  \((\frac{1}{6},{\bf 2},  {\bf 3}) \)  \\ 
\(\hat{d}^{3}\) &\Yes &-& \((\frac{1}{3},
{\bf 1},  {\bf \overline{3}}) \)  \\ 
\(\hat{u}^{3}\) &\Yes &-& \((-\frac{2}{3},
{\bf 1},  {\bf \overline{3}}) \)  \\
\(\hat{l}^{3}\) &\Yes &-&\((-\frac{1}{2}, {\bf 2}, {\bf 1}) \) \\
\(\hat{e}^{3}\)&\Yes &-& \((1, {\bf 1},
 {\bf 1}) \)  \\  \hline\hline
\(\hat{H}_d\)   & - &\Yes  & \( (-\frac{1}{2},{\bf 2},  {\bf 1}) \) \\ 
\(\hat{H}_u\) &  - &\Yes      & \( (\frac{1}{2},{\bf 2},  {\bf 1}) \)  \\ 
 \hline
\(\hat{F}_-\)   & - &\Yes     & \( (-1,{\bf 1},  {\bf 1}) \) \\ 
\(\hat{F}_+\) & - &\Yes       & \( (1,{\bf 1},  {\bf 1}) \)  \\ 
 \hline\hline
\(\hat{B}_V\)   & - &\Yes     & \( (0,{\bf 1},  {\bf 1}) \) \\ 
\(\hat{W}_V\) & - &\Yes       & \( (0,{\bf 3},  {\bf 1}) \)  \\ 
\(\hat{G}_V\)   & - &\Yes     & \( (0,{\bf 1},  {\bf 8}) \) \\ 
 \hline\hline
\end{tabular} \caption{The matter content of model 2. 
\label{Table:Model2}}
\end{center} 
\end{table}

\par  The superpotential for both models is given by 
\begin{align} 
W = & \,  Y_u\,\hat{u}\,\epsilon_{ij} \hat{q}^i\,\hat{H}^j_u\,- Y_d
\,\hat{d}\,\epsilon_{ij} \hat{q}^i\,\hat{H}^j_d\,- Y_e \,\hat{e}\,\epsilon_{ij}
\hat{l}^i\,\hat{H}^j_d +\mu H_u H_d + \acute{\mu} F^{-} F^{+}\,\,.
\end{align} \label{eq:MSSM}
It would be very worthwhile to consider the generation of the term $ \acute{\mu} F^{-} F^{+} $  in the superpotential, although for this paper we will not need to consider it, and postpone that to later work.  We will now explore the running parameters of these two theories as one changes the scale of the extra dimension.

\section{Running parameters}\label{sec:running}
\begin{figure}[!th]
\begin{center}
\includegraphics[width=7.5cm,angle=0]{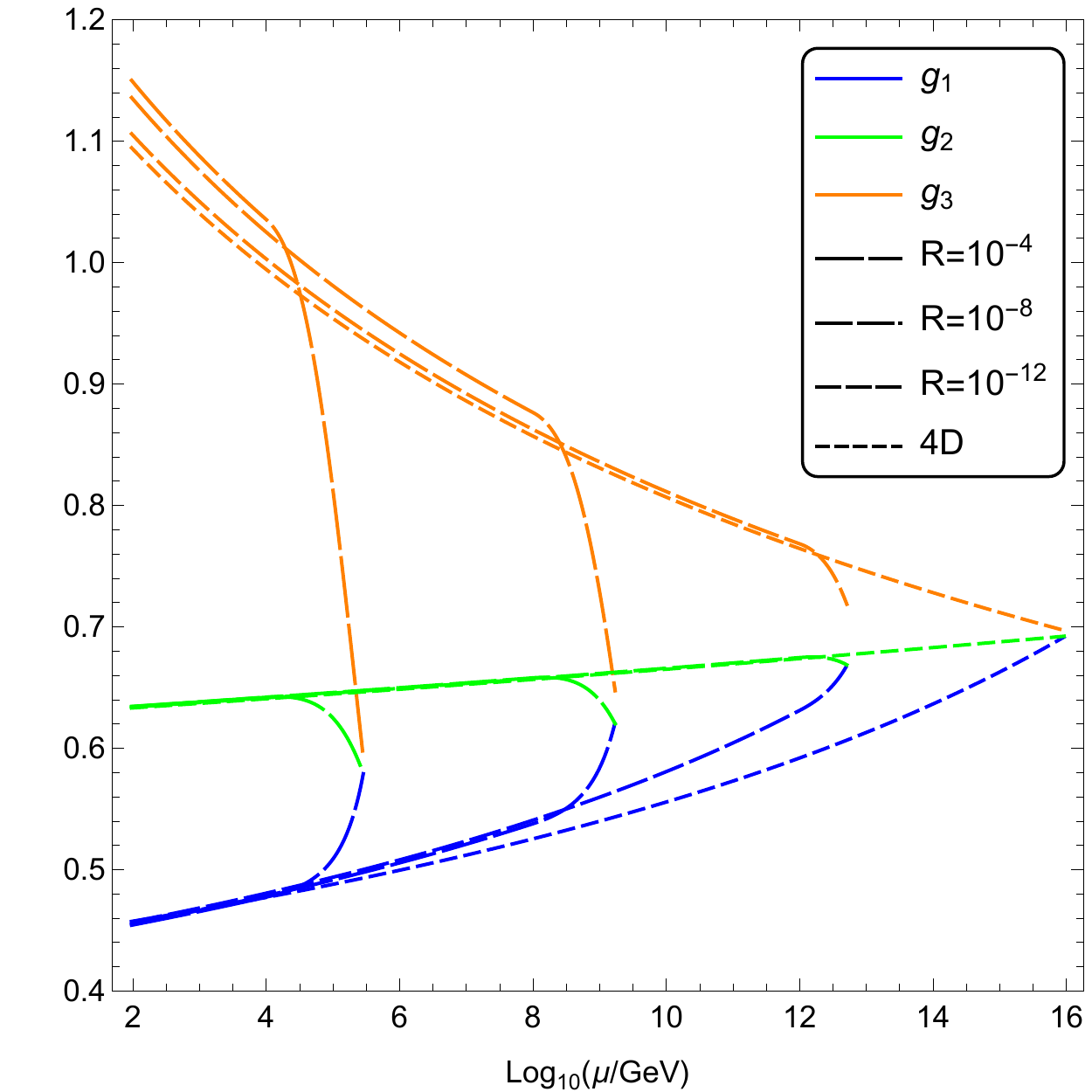}
\includegraphics[width=7.5cm,angle=0]{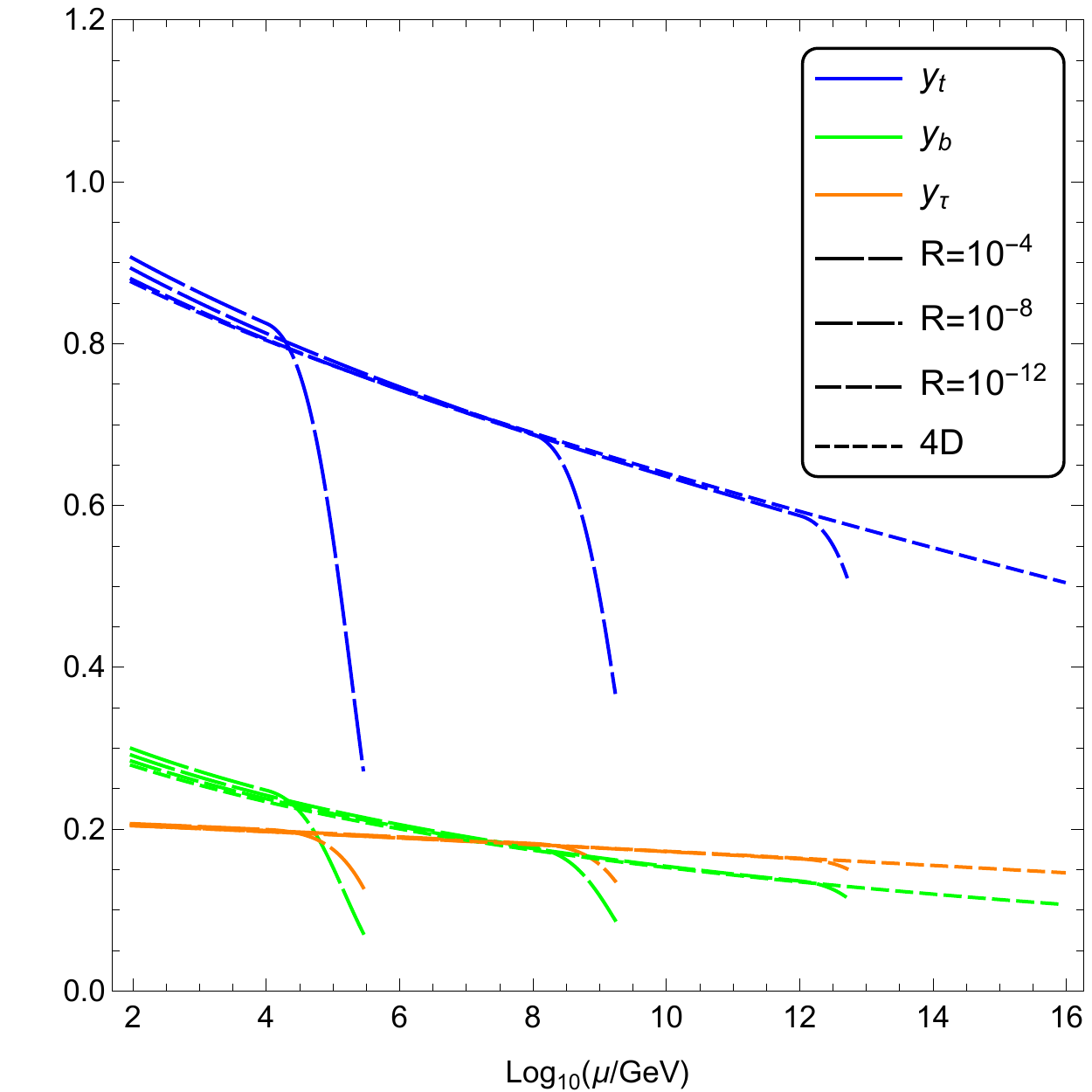}\qquad
\includegraphics[width=7.5cm,angle=0]{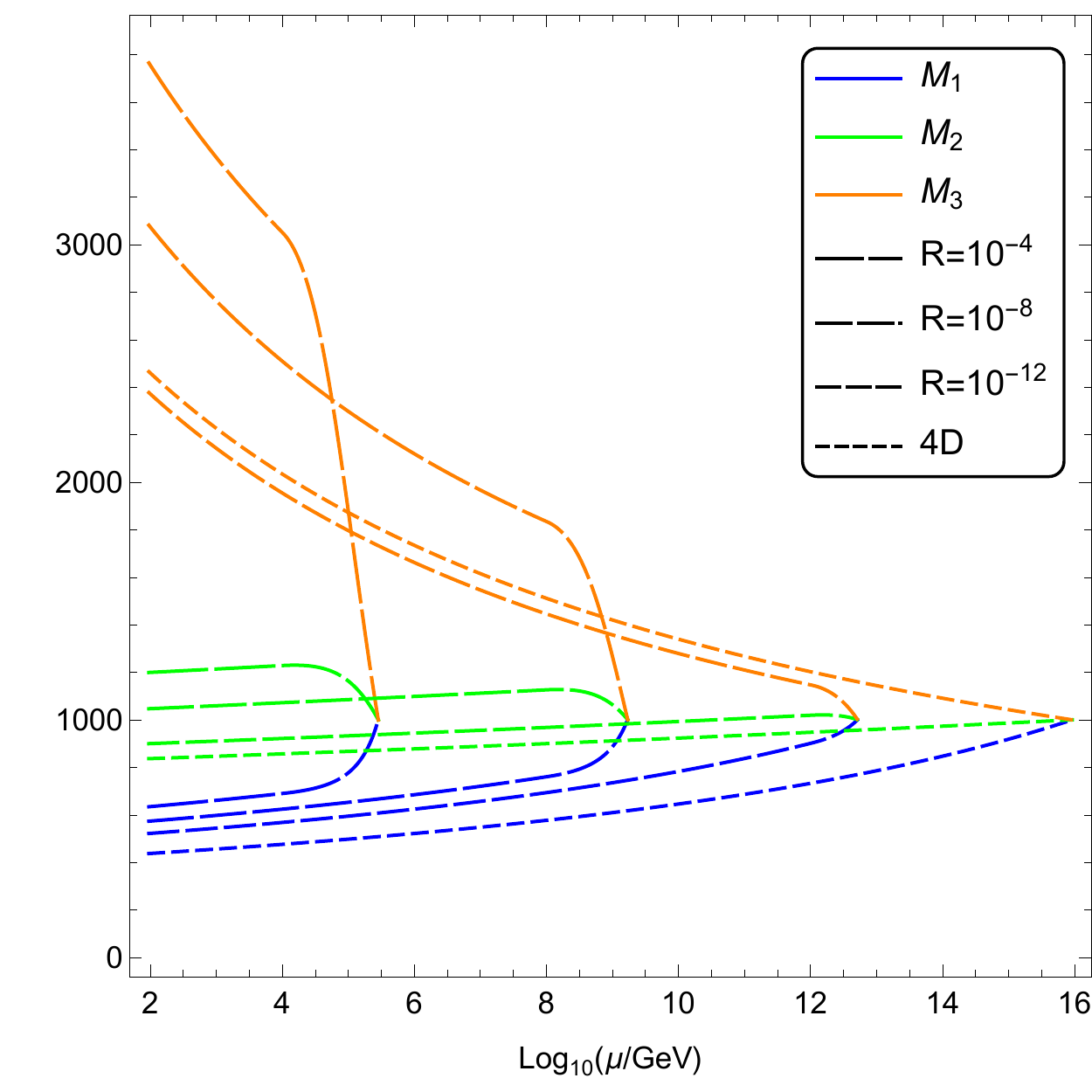}
\includegraphics[width=7.5cm,angle=0]{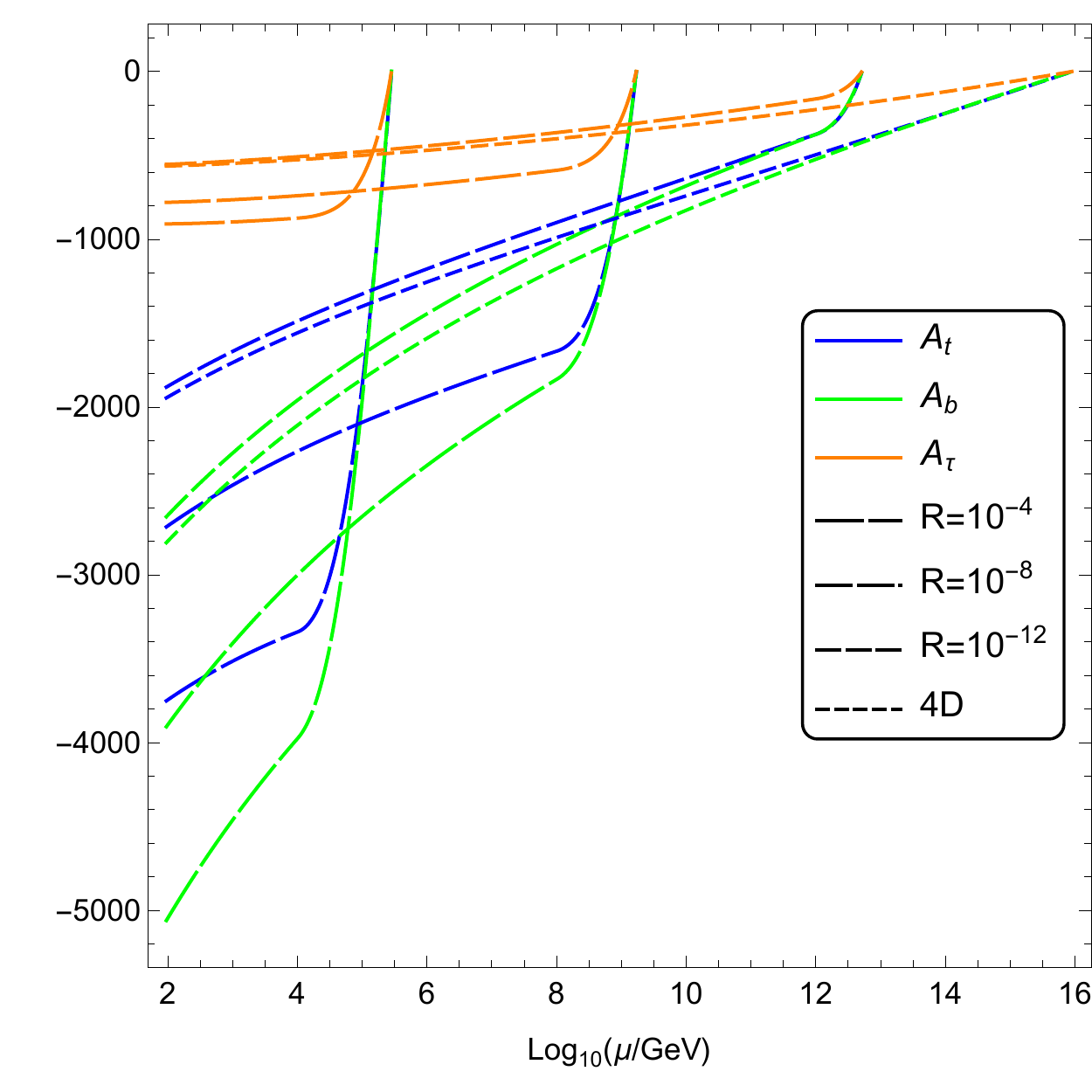}
\caption{{\it Model 1 running of gauge coupling constants $g_i(\mu)$ (top left panel), 3rd generation Yukawa couplings (top right panel), trilinear soft terms (bottom right panel) and gaugino soft terms (bottom left panel) with compactification scales $1/R\sim$ $10^4$\gev,  $10^8$\gev \& $10^{12}$\gev,  as a function of  $Log_{10}$($\mu$/\gev). In this example all soft terms were set to $M_{SUSY}=1$\tev} at the unification scale (defined by $g_1=g_2$), except the trilinear soft terms ($A_i$) which were set to $0$.} 
\label{fig:running1}
\end{center}
\end{figure}

It is particularly interesting to understand and compare the behaviour of the various running parameters of these theories compared to the more usual four dimensional MSSM. The behaviour of the various parameters as a function of renormalisation scale for model 1 is pictured in figure \ref{fig:running1}. Of particular note is that unification happens much earlier if the size of the extra dimension is large \cite{Dienes:1998vg} , than the usual four dimensional case.  One also finds that the top Yukawa reduces rather significantly and becomes of similar order to the other Yukawa couplings near the unification scale.   In addition one finds that even for initially vanishing A-terms the $A_t$ term may become multi-TeV in value at the electoweak scale, which is encouraging from the perspective of obtaining the observed $125 \gev$ Higgs mass.  It is also the case (bottom left) that the gluino mass can become much hearvier than the other gauginos allowing for the theory to still have a light bino and wino whilst allowing for a gluino above current exclusions.

\begin{figure}[!th]
\begin{center}
\includegraphics[width=7.5cm,angle=0]{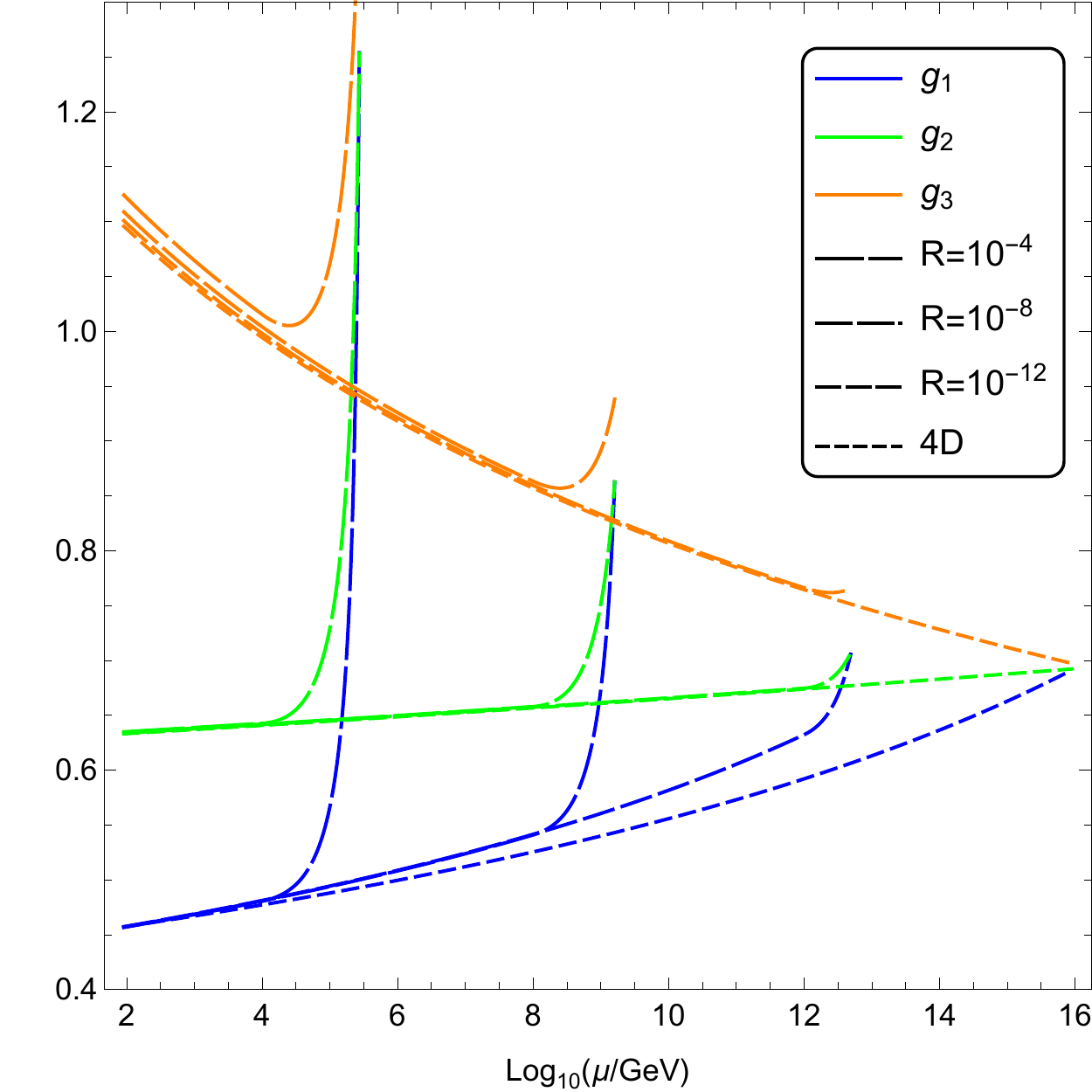}
\includegraphics[width=7.5cm,angle=0]{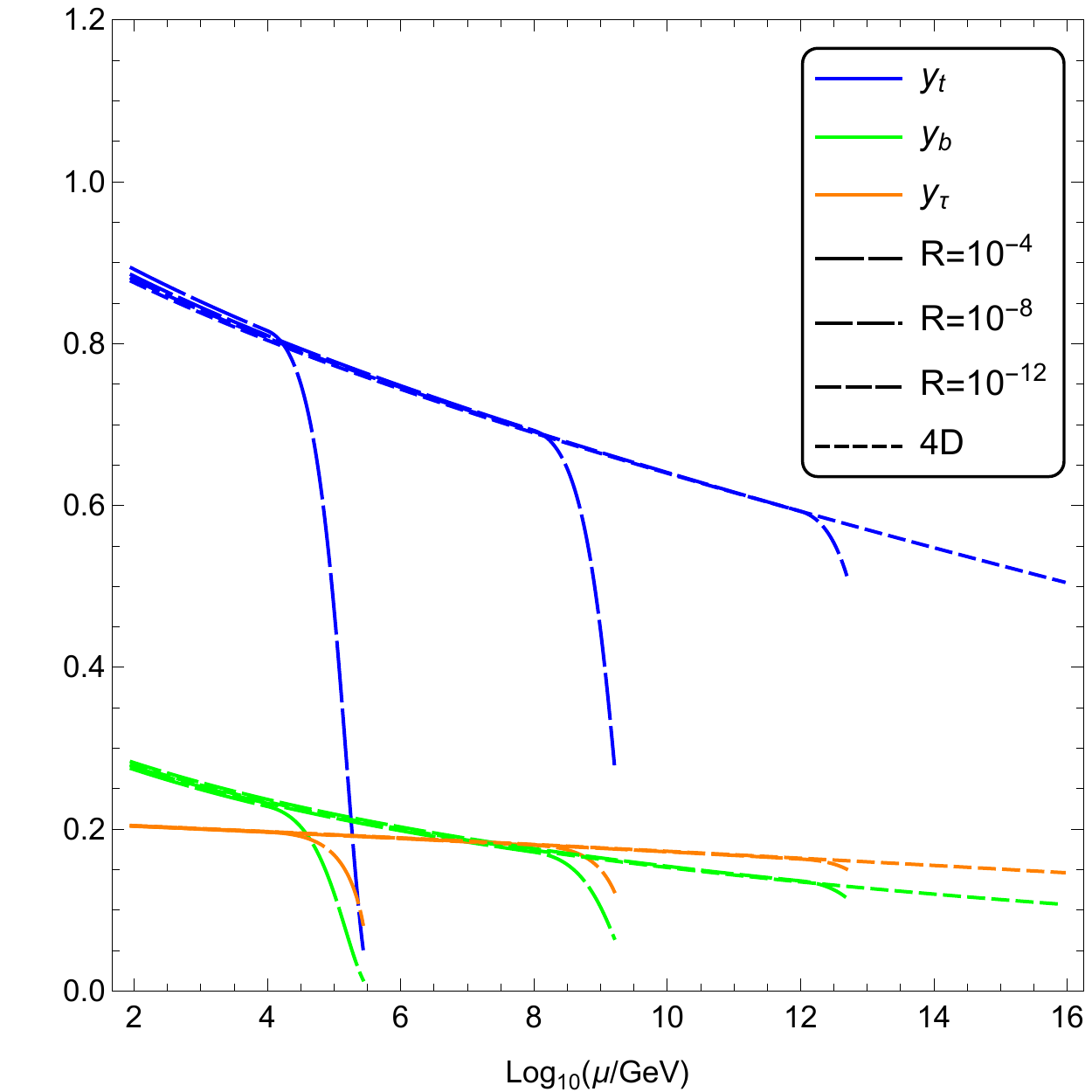}\qquad
\includegraphics[width=7.5cm,angle=0]{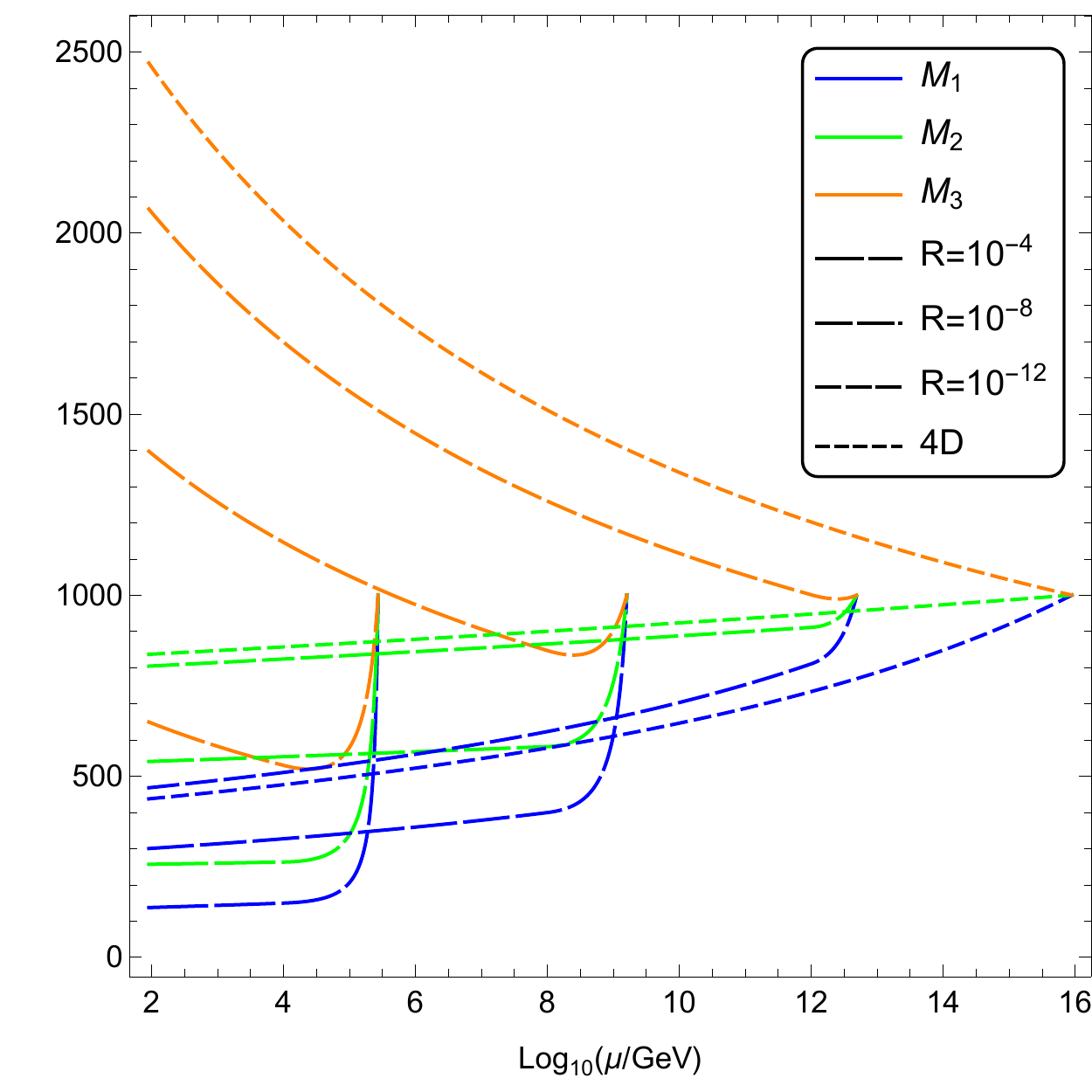}
\includegraphics[width=7.5cm,angle=0]{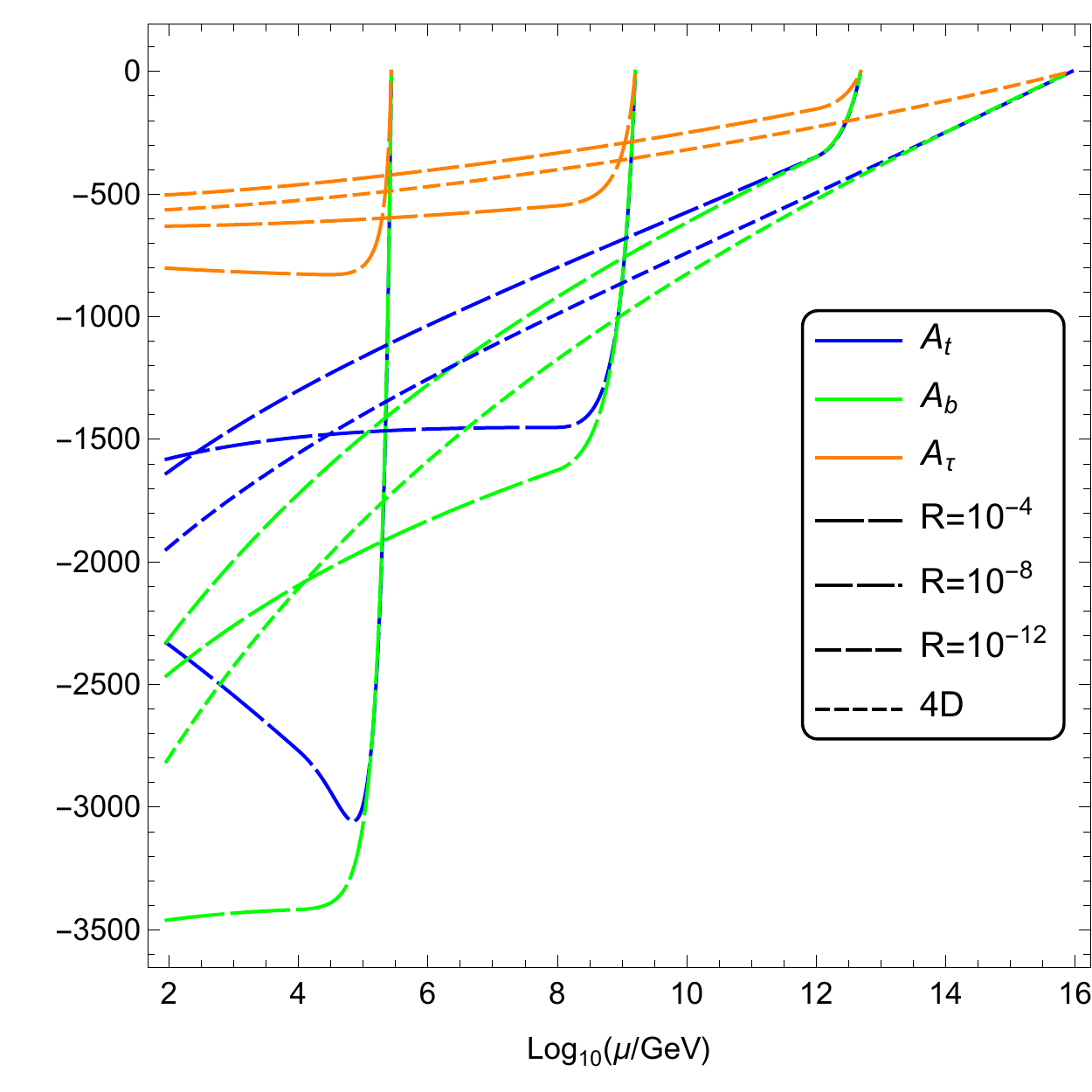}
\caption{{\it Model 2 running of gauge coupling constants $g_i(\mu)$ (top left panel), 3rd generation Yukawa couplings (top right panel), trilinear soft terms (bottom right panel) and gaugino soft terms (bottom left panel) with compactification scales $1/R\sim$ $10^4$\gev,  $10^8$\gev \& $10^{12}$\gev,  as a function of  $Log_{10}$($\mu$/\gev). In this example all soft terms were set to $M_{SUSY}=1$\tev} at the unification scale (defined by $g_1=g_2$), except the trilinear soft terms ($A_i$) which were set to $0$.} 
\label{fig:running2}
\end{center}
\end{figure}

The first model may be compared with model 2 similarly presented in figure \ref{fig:running2} and in table \ref{Table:Model2}.  In these figures it is notable that  that gauge couplings quite nearly unify but the gauge couplings rise rather than fall, after the KK modes start to take effect in the RGEs. The $Y_t$ still decreases in value, although now rather interestingly the $A_t$ becomes so quickly negative that it can quickly overcompensate the effect of the gluino soft mass, and for very large radius, the $A_t$ running may even return on itself.  Again the wino and bino soft terms can be much smaller than that of the gluino, even starting from the same initial value.


\section{Supersymmetry breaking in benchmark models}\label{sec:benchsusy}
So far our exploration has been reasonably agnostic about how supersymmetry is broken, since the main feature of the models presented in the previous sections are their RGEs. In what follows we will simply refer to sets of RGEs we used, as models.  

There are however a number of ways that have been proposed for the parametrisation of supersymmetry breaking in a five dimensional scenario. In this section we wish to identify these scenarios and look at their patterns of supersymmetry breaking which define their possible high scale spectra.
\subsection{Gravity mediation (CMSSM)}
Our first benchmark scenario is the simple CMSSM spectrum, however since easier generation of $A$-terms during running is a key feature of five dimensional running, we will always take $A_i=0$ case for which the difference between five and four dimensional theories is the most visible. This implies a very simple type of spectrum with just two free parameters $M_{\frac{1}{2}}$ and $m_0$: 
\be
M_{i}=M_{\frac{1}{2}} \ \  ,  \ \  m^2_{\tilde{f}}=m^2_0 \ \ 
, \ \ A_i=0,
\ee 
defined at the unification scale.
\subsection{Minimal Gauge Mediation (MGM) in five dimensions}
In gauge mediation there is an additional characteristic scale at which SUSY is broken, which for brevity we will labelled $M$. For the five dimensional RGEs to have an impact on the spectrum and to not simply be an effective four dimensional theory with a low SUSY breaking scale we wish that M is at least $O(1/R)$ and possibly nearer $M_{\text{unification}}$.

The soft terms in five dimensional GMSB, at the breaking scale, are then given by 
\be\label{eq:MGMsoftterms}
M_{r}=\left(\frac{\alpha_r}{4\pi}\right)\left(\frac{F}{M}\right) \ \  ,  \ \  m^2_{\tilde{f}}\simeq2\sum_r C^{r}_{\tilde{f}}\left(\frac{\alpha_r}{4\pi}\right)^2 \left(\frac{F}{M}\right)^2\left( \frac{1}{M R}\right)^2  \ \ 
, \ \ A_i=0,
\ee
where $F$ and $M$ are the free parameters we will scan over.   This paper is the first implementation of five dimensional GMSB soft masses \cite{Kaplan:1999ac,Chacko:1999mi,Mirabelli:1997aj,McGarrie:2010kh,McGarrie:2011av}, with five dimensional RGEs  \cite{Abdalgabar:2015ora,Abdalgabar:2014bfa}.  In  both  model 1 and 2 we will take the supersymmetry breaking to be on the opposite brane to the matter, and both brane and bulk matter are essentially suppressed by the effect of the extra dimension, as in the above equation.
\subsection{Realising natural SUSY with GMSB in five dimensions (nMGM)}
\label{sec:natural}
\begin{figure}[!th]
\begin{center}
\includegraphics[width=4cm,angle=0]{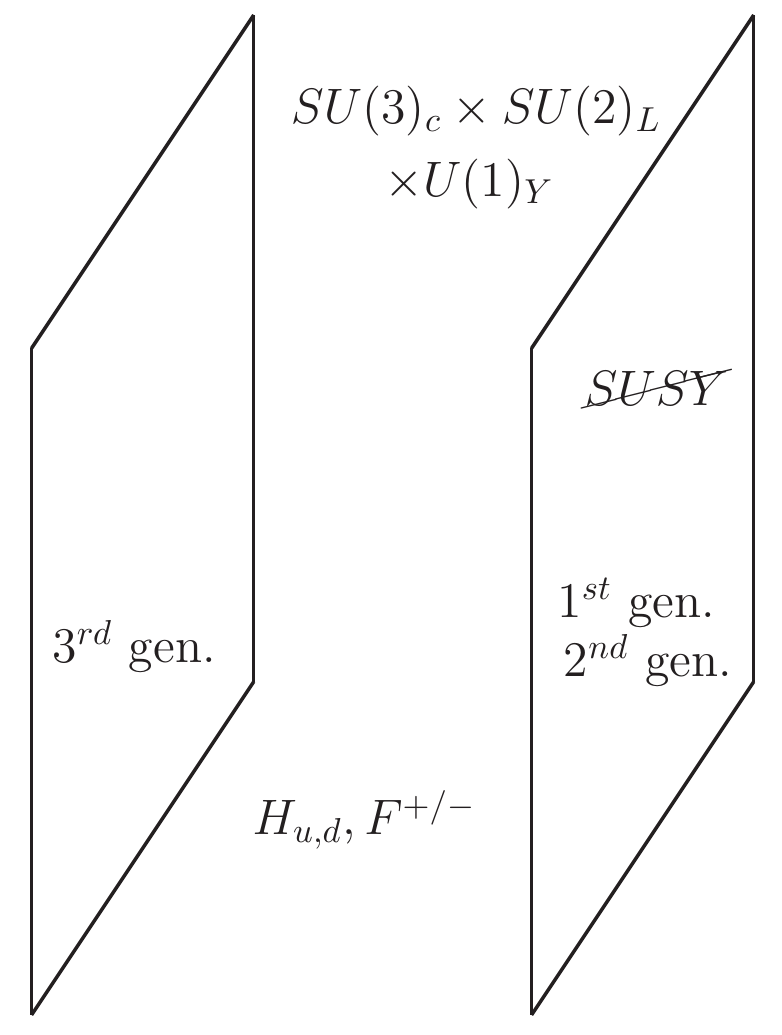}
\caption{{\it 
Pictorial to represent the location of matter in the five dimensional model resulting in a natural SUSY breaking scenario with GMSB (nMGM spectrum) and model 1 RGEs.
}}  
\label{fig:naturalmodel}
\end{center}
\end{figure}

The renormalisation group equations of model 1 may be used to explore a natural susy scenario as pictured in figure \ref{fig:naturalmodel}. In this model the 3rd generation is located on one brane and the 1st and 2nd generation on another, along with the supersymmetry breaking sector. The effects of supersymmetry breaking are mediated by gauge forces \cite{Giudice:1998bp} (but one can also easily consider gravity mediation too in this context) and the result is that the 1st and 2nd generation and also the gauginos will receive normal (4D) GMSB soft mass contributions but the 3rd generation will be heavily suppressed \cite{McGarrie:2011av,McGarrie:2010kh,Abdalgabar:2014bfa,Abdalgabar:2015ora}.  The soft mass matrix for squarks and sleptons takes the form 

\be
m_{\tilde{f}}^2(M_{\cancel{SUSY}}) \sim \Lambda^2 \left(\begin{array}{ccc}
1& 0 & 0 \\
 0 & 1& 0 \\
 0 & 0 & 0\\
\end{array}\right)+...\label{eq:softheirarchy}
\ee
leading to an interesting natural SUSY spectrum of lighter 3rd generation squarks. This scenario suggests that natural SUSY softer terms are imprinted due to the `geometry' of the theory. We will consider such a natural spectrum in context of minimal gauge mediation, the resulting soft terms are similar to those in\eqref{eq:MGMsoftterms}, however now only third generation sfermions are suppressed by $1/(MR)^2$. In the text we will refer to this as an nMGM spectrum.  Needless to say, a similar model may be constructed using brane to brane gravity mediation.
It would also be interesting to discuss models with $H_u$ and $H_d$ localised alongside the 3rd families, however it would require a much more serious modification of the RGE's of our Model 1 and 2, consequently we postpone that discussion to future work. 
\section{Electroweak symmetry breaking and naturalness}\label{sec:ewsb}
\begin{figure}[!th]
\begin{center}
\includegraphics[width=6.2cm,angle=0]{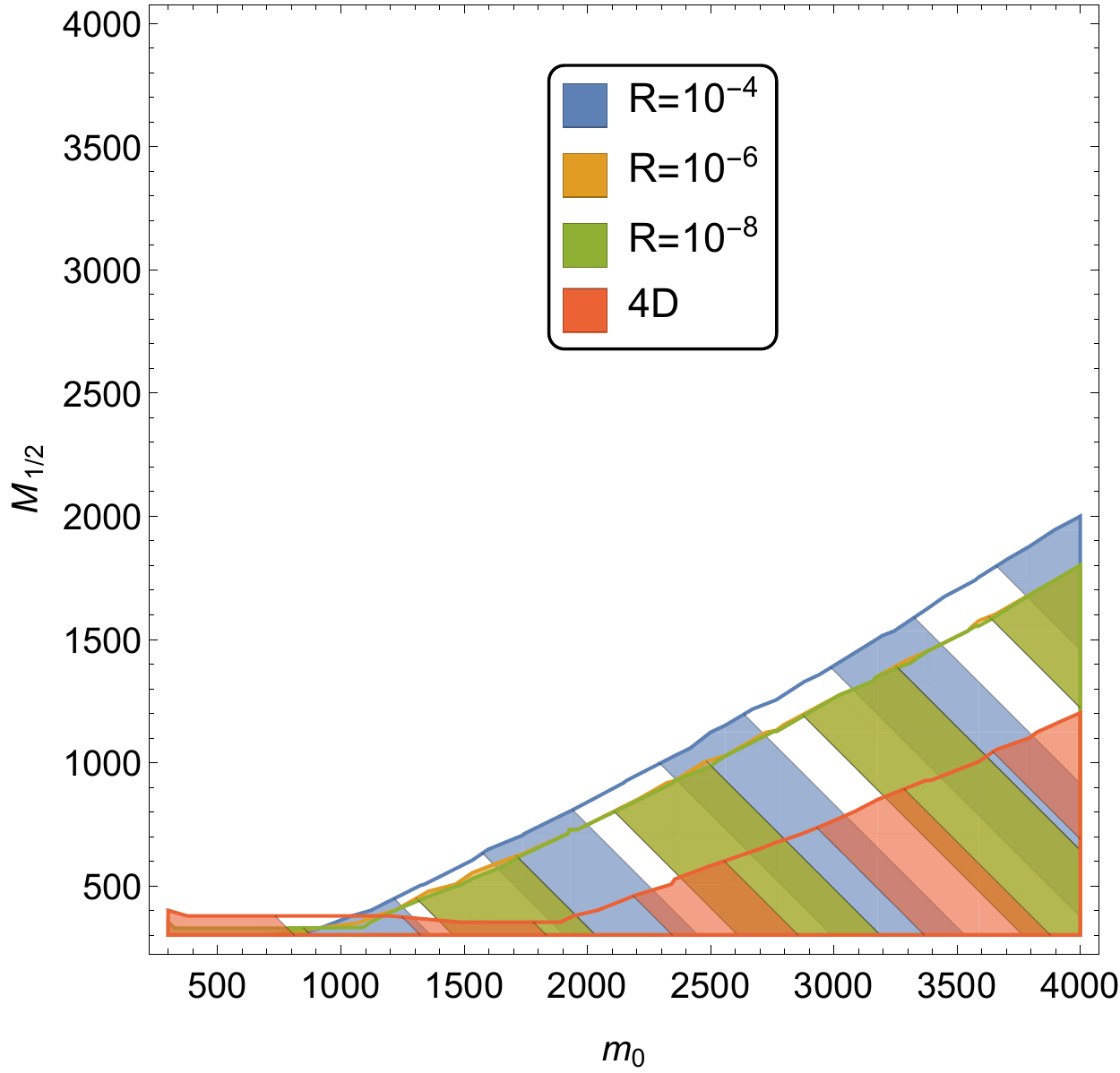}
\includegraphics[width=6.2cm,angle=0]{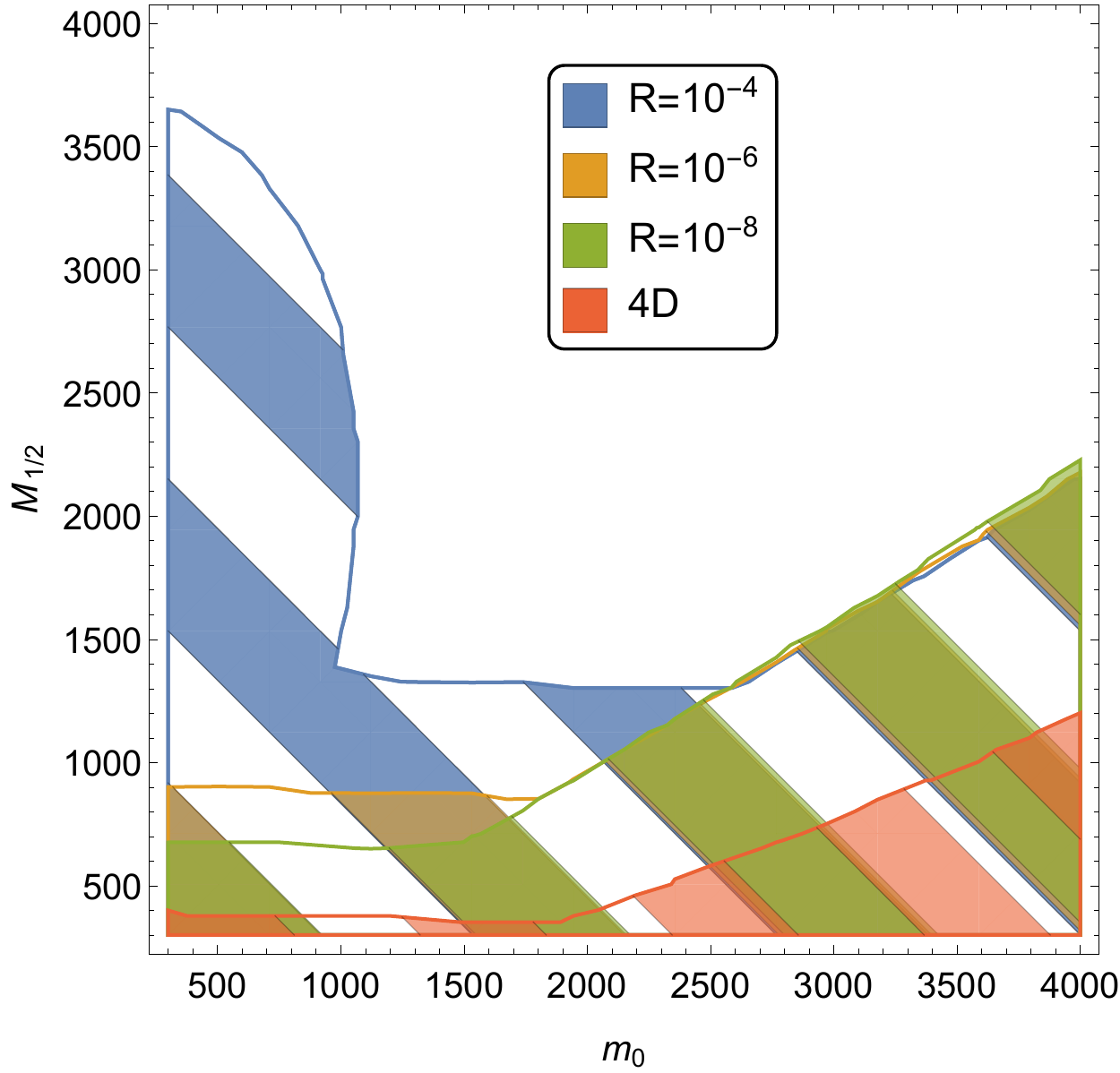}
\includegraphics[width=6.2cm,angle=0]{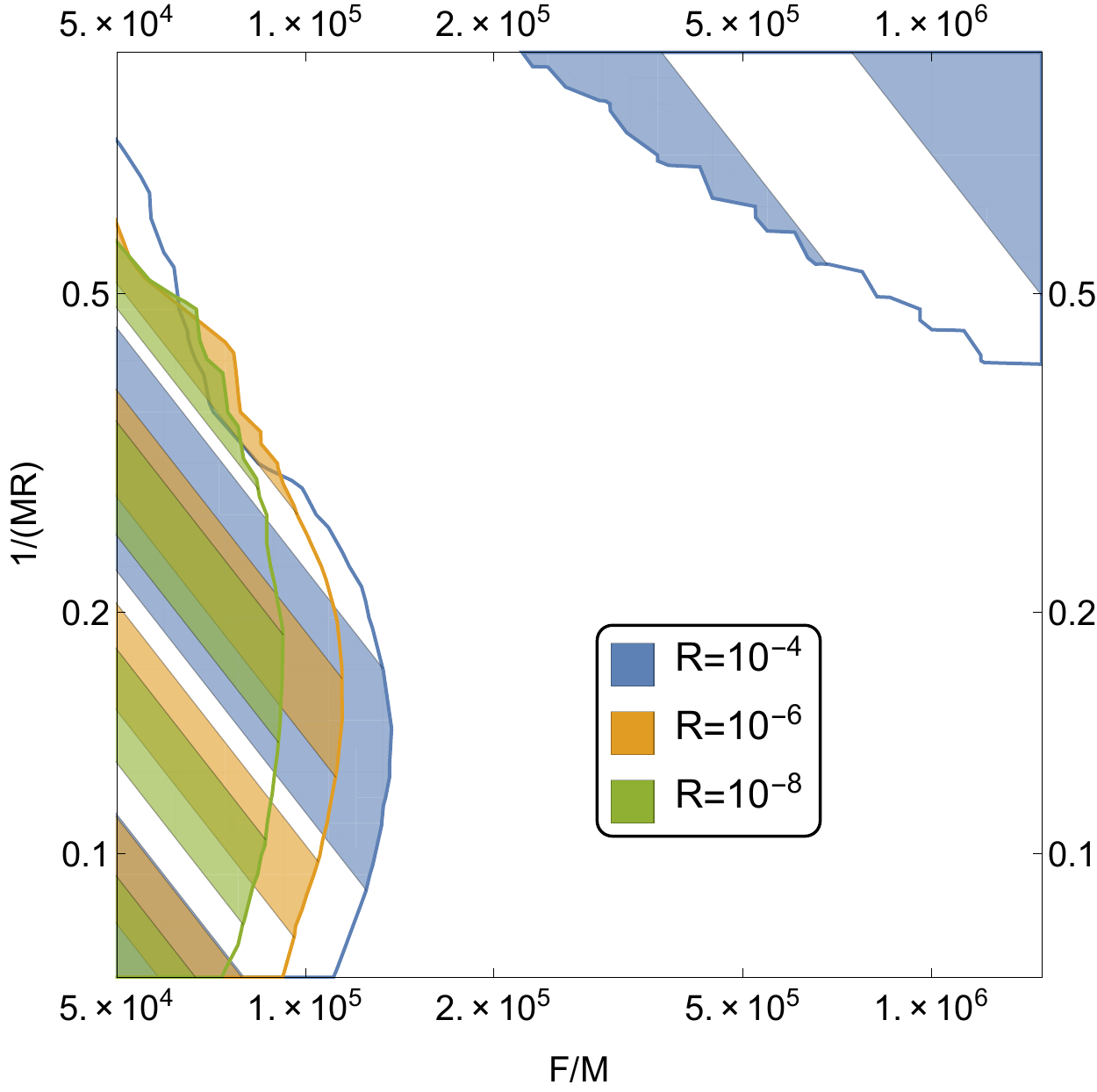}
\includegraphics[width=6.2cm,angle=0]{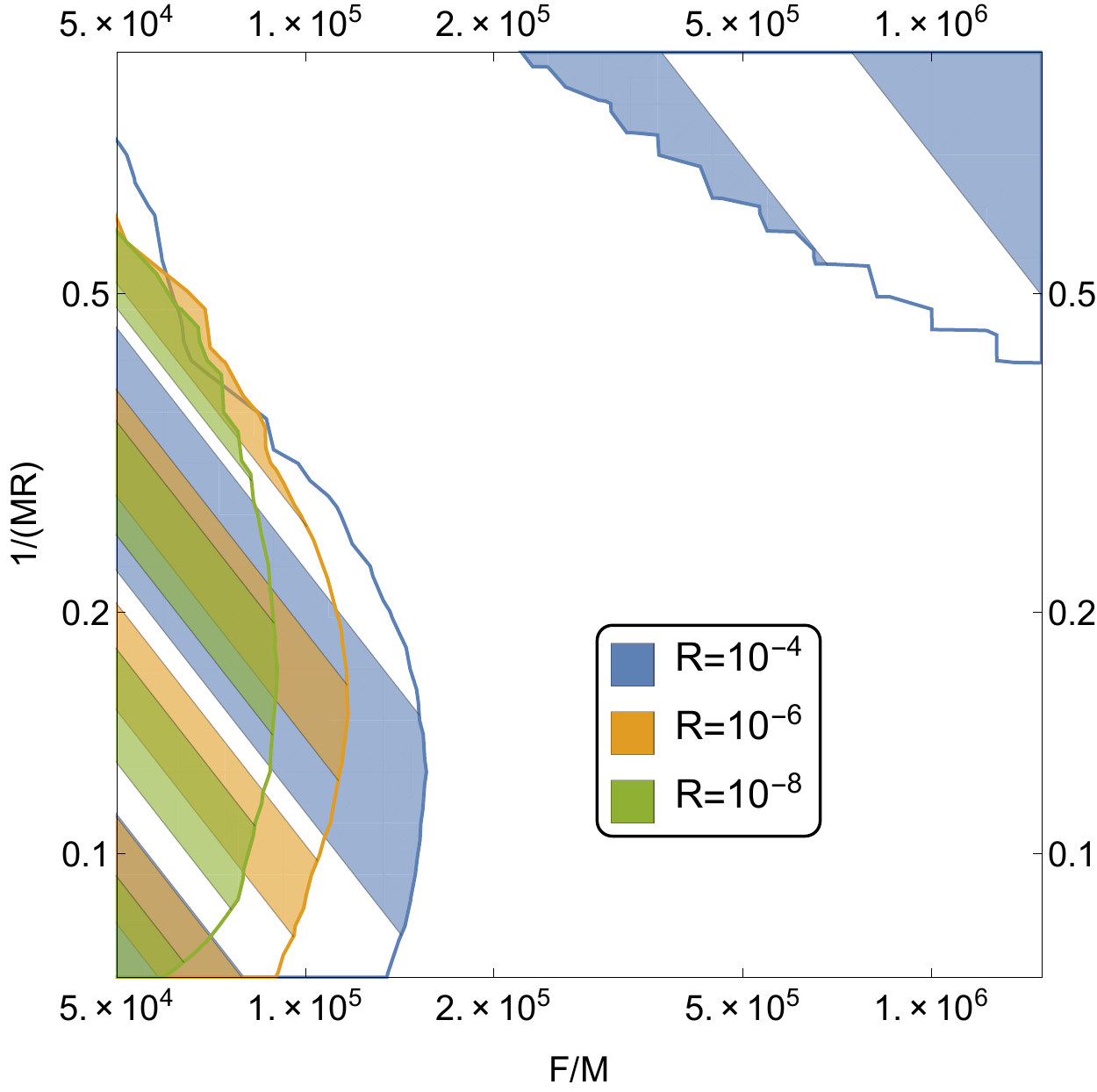}
\includegraphics[width=6.2cm,angle=0]{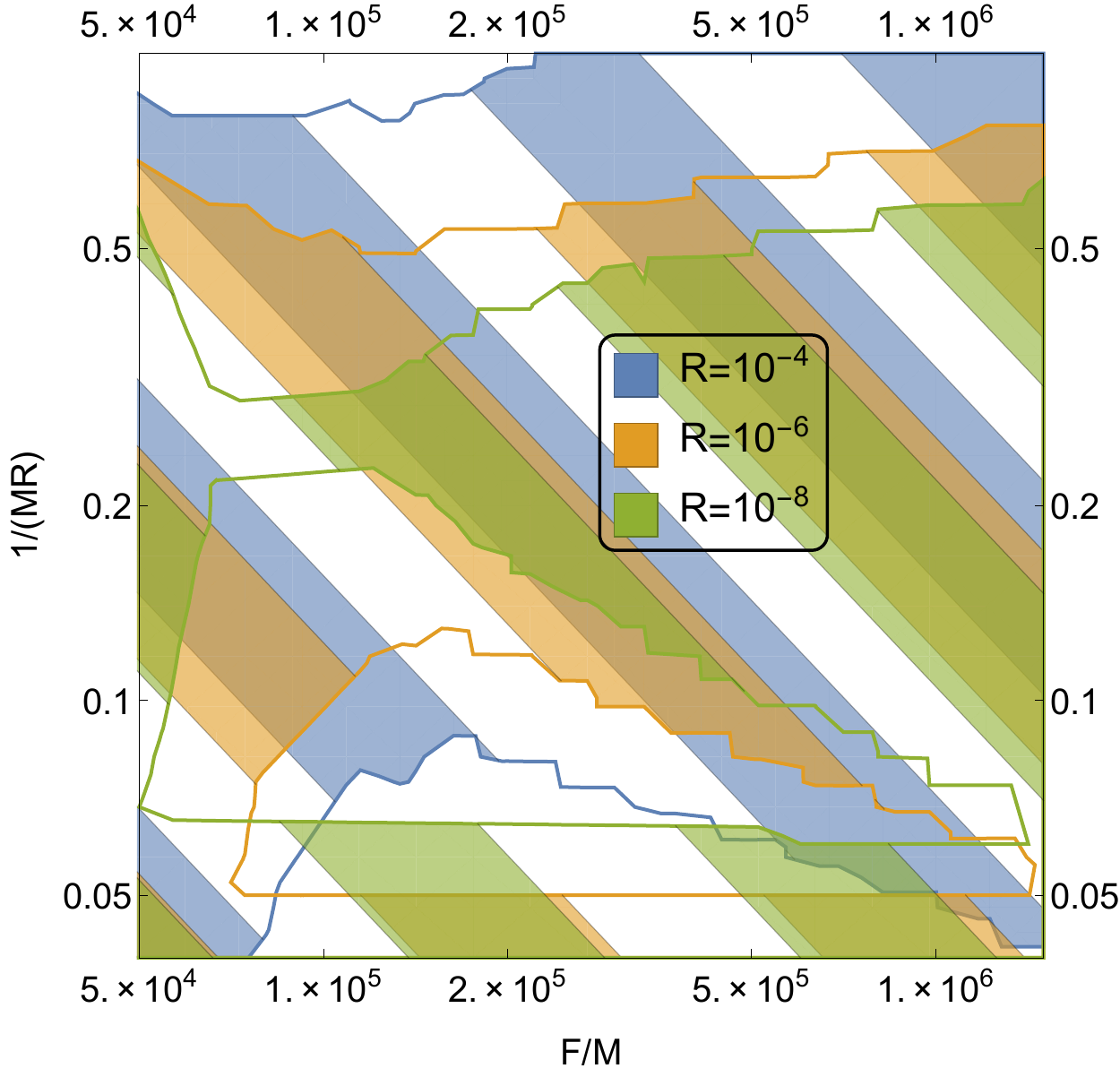}
\hspace{6.2cm}
\caption{{\it 
Striped regions of the CMSSM (top row) MGM (middle row) and nMGM (bottom row) parameter space cannot accommodate electroweak symmetry breaking or are already excluded by direct searches. Left hand side plot shows results for model 1 and right hand side for model 2.  Both show different sizes of the extra dimension and CMSSM shows the $4D$ case as well.
}}  
\label{fig:EWSBplot}
\end{center}
\end{figure}

One important feature of a model is whether its parameter space can accommodate electroweak symmetry breaking. Figure~\ref{fig:EWSBplot} shows regions in the parameter space of our models where the breaking does not occur or which violate direct detection bounds summarised in table \ref{tab:expexcl} \cite{Agashe:2014kda}. Exclusions corresponding to varying size of the extra dimension (including the $4D$ case) are plotted together.
\begin{table}
\begin{center} 
\begin{tabular}{c|c} 
particle & mass bound in \gev \\
\hline
$\tilde{g}$ & $1200$  \\
$\tilde{q}_{1,2}$ & $800$ \\
$\tilde{t}$ & $700$  \\
$\tilde{b}$ & $650$  \\
$\tilde{\chi}^{\pm}_1$ & $92$ \\
$\tilde{\chi}^{0}_1$ & $46$ \\
\end{tabular}
\end{center}
\caption{{\it 
Experimental exclusion limits used
\label{tab:expexcl}
}} 
\end{table}
For standard CMSSM and MGM boundary conditions Model $1$ predicts rather standard spectra of sparticles quite similar to the $4D$ case. However Model 2 due to much lower gaugino masses compared to the $A$-terms allows us to obtain very light stops and maximal mixing even despite $A$-terms vanishing at the unification scale. In fact for large $R=10^{-4}$ the peculiar shape of the CMSSM excluded region in model 2 comes from obtaining too light stops that would have already been observed. 

The MGM excluded region comes from the interplay between large scalar masses we obtain at the scale $M$ when $M=1/R$, and when they are generated during 5D modified running between $1/R$ and $M>>1/R$. The minimal stop mass is obtained between these two situations and results in excluded part on the left hand side of middle row in Figure~\ref{fig:EWSBplot} where the small stop soft mass fails to push $m_{H_u}$ to negative values and break electroweak symmetry.

This is also visible in nMGM plot on the bottom row of Figure~\ref{fig:EWSBplot}. However here the problem is more severe since $m_{H_u}$ is not suppressed by $1/(MR)$ at the SUSY breaking scale, and a much bigger part of the parameter space is excluded. 
For nMGM spectrum this problem appears also for very small $1/(MR)$, because in this part of the parameter space the difference between Higgs and stop soft masses is the largest. These two effects lead to appearance of a window of allowed parameter space which is very interesting, since it is in that window, that we obtain the highest Higgs mass.

\subsection{Exploring naturalness in benchmark scenarios}\label{sec:benchnatural}
In MSSM-like theories, at a finite loop order, electroweak symmetry breaking is radiatively induced. The up-Higgs soft mass is driven to negative values, leading to
\be
-\frac{1}{2}M_z^2= m^2_{H_u}(\Lambda) + \delta m_{H_u}^2 + |\mu|^2  +\mathcal{O}\left(\tan \beta^{-2}\right)
\ee
At leading order, the running of this soft mass in four dimensions follows
\be
 \delta m_{H_u}^2  \sim - \frac{3}{8\pi^2}y_t^2 (m_{Q_3}^2+m_{U_3}^2 +A_t^2)\text{Log}\left(\frac{M_{SUSY}}{M_S}\right)
\ee
In five dimensional models the RGEs are rather different due to the power law contributions and one finds
\be
 \delta m_{H_u}^2  \sim - \frac{3}{8\pi^2}y_t^2 (m_{Q_3}^2+m_{U_3}^2 +A_t^2)\left[\text{Log}\left(\frac{1}{R M_S}\right)+M_{SUSY}R \right]
\ee

\begin{figure}[!th]
\begin{center}
\includegraphics[width=6.2cm]{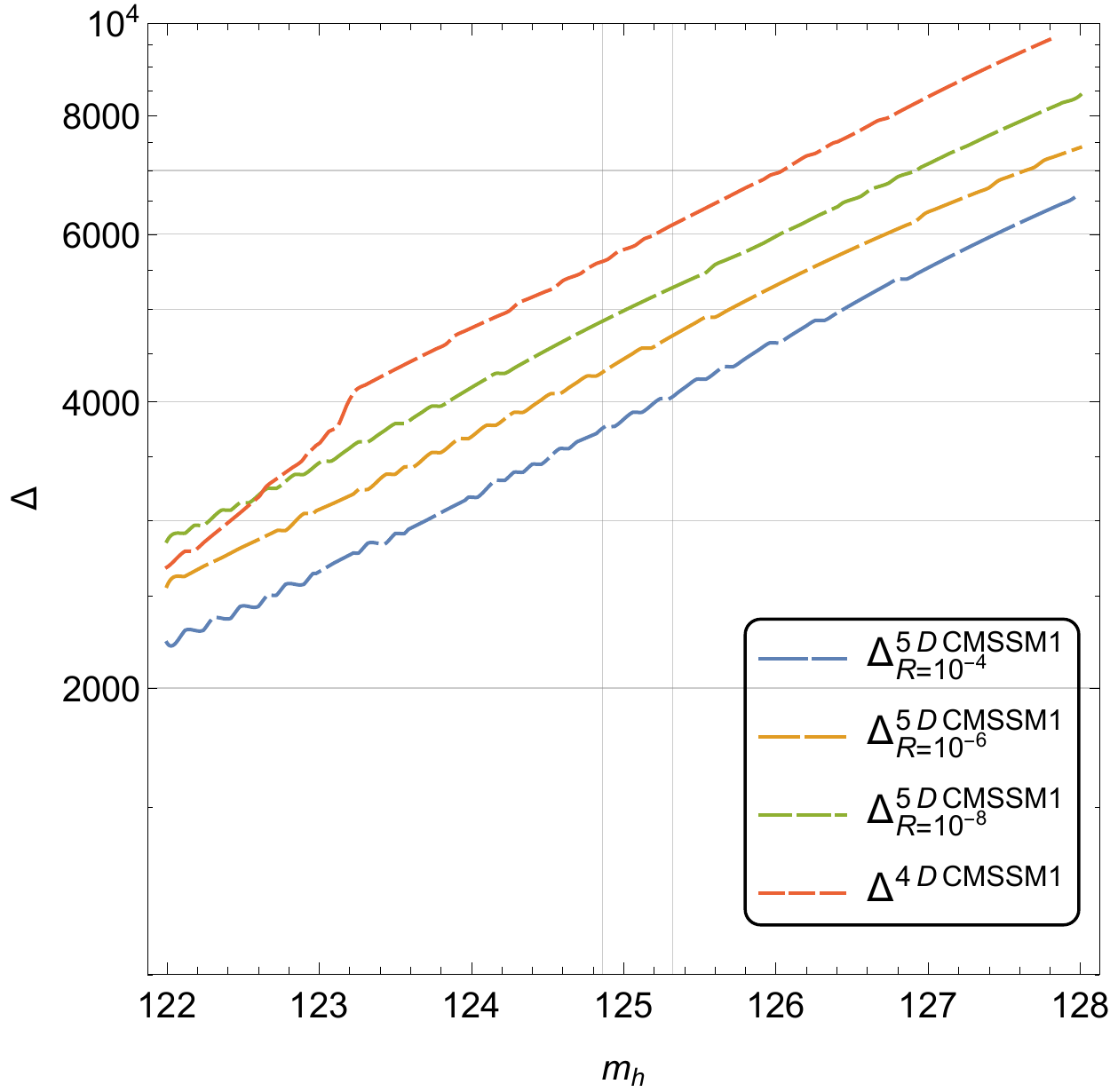}
\includegraphics[width=6.2cm]{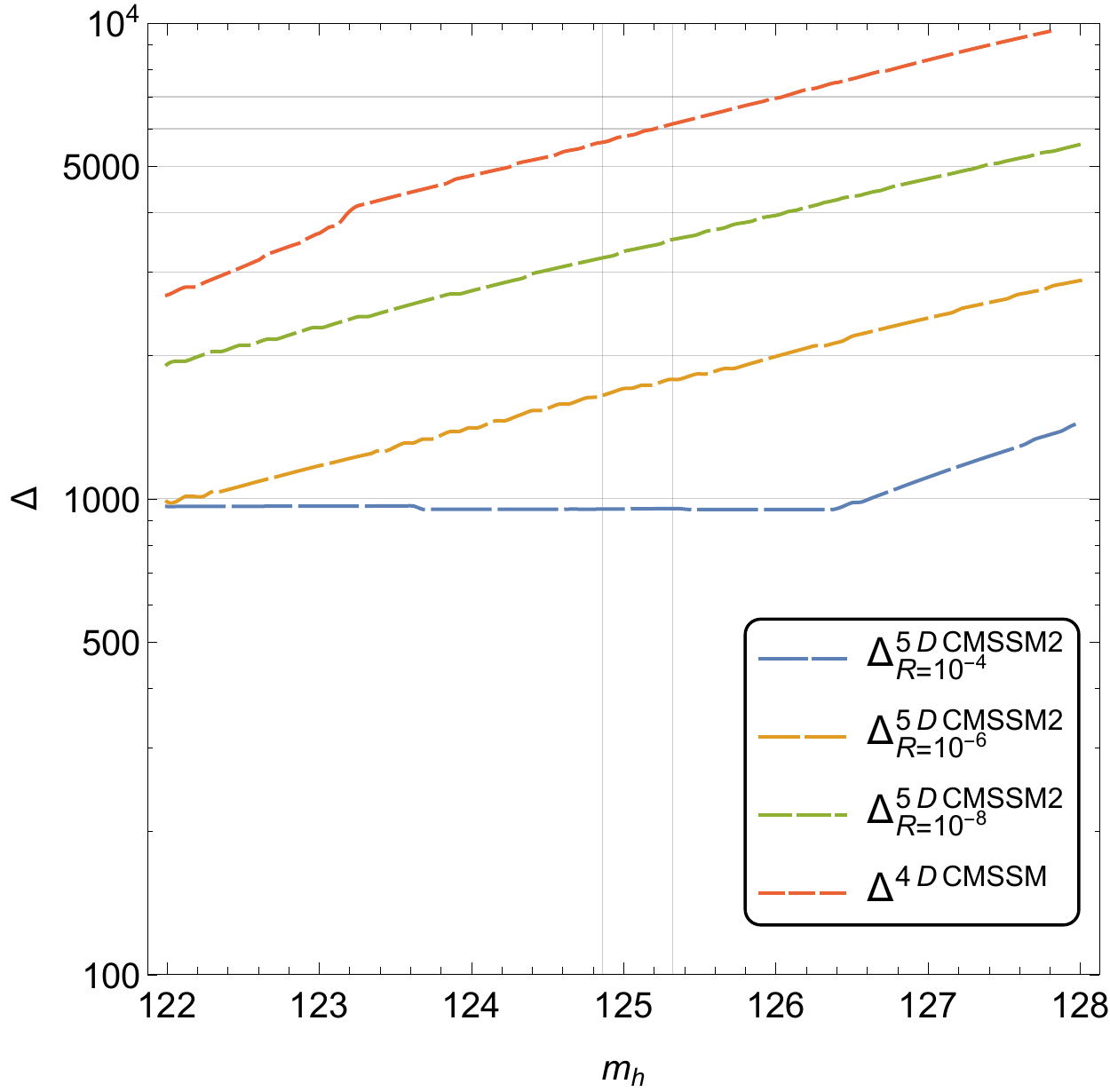}
\includegraphics[width=6.2cm]{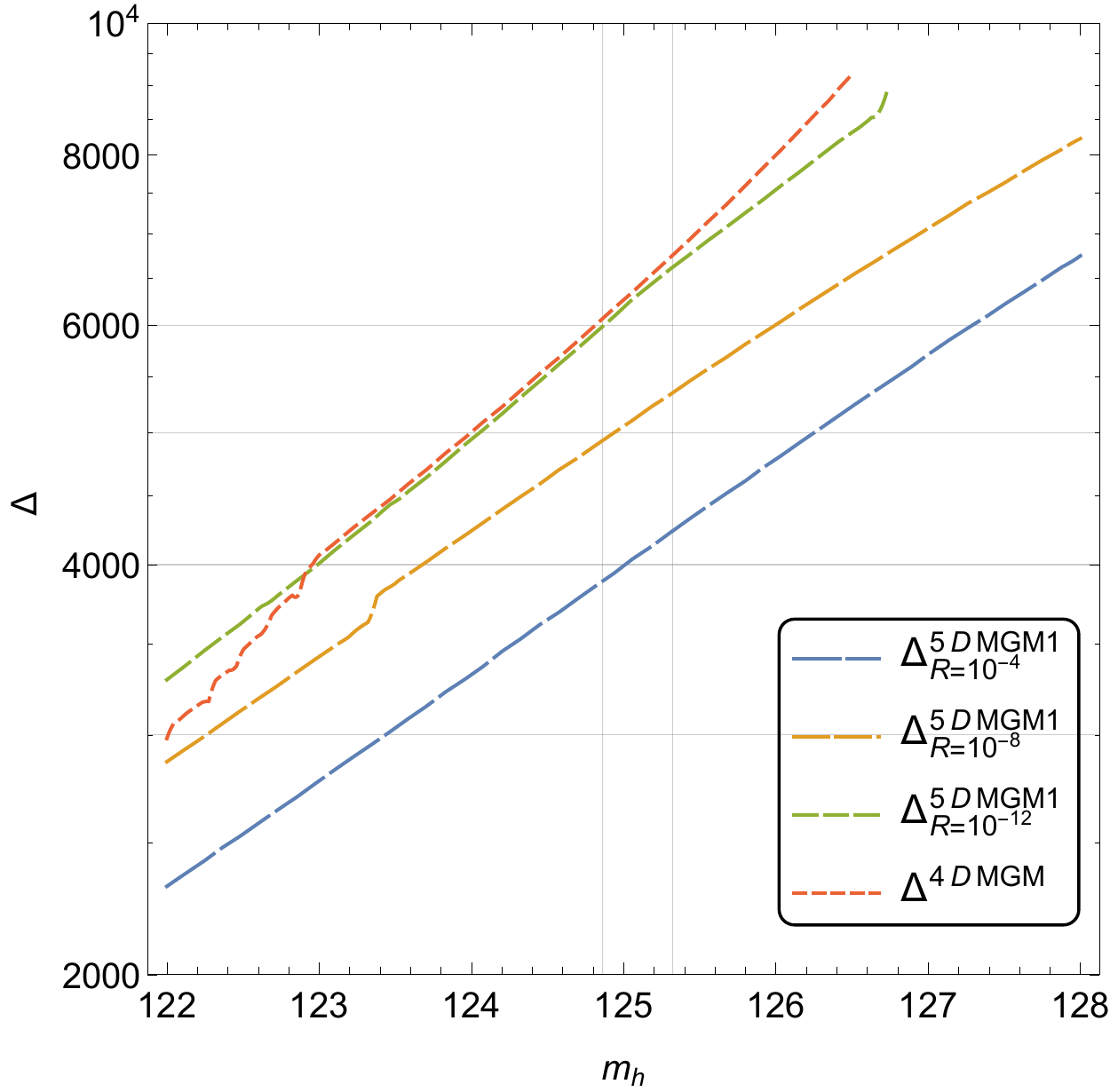}
\includegraphics[width=6.2cm]{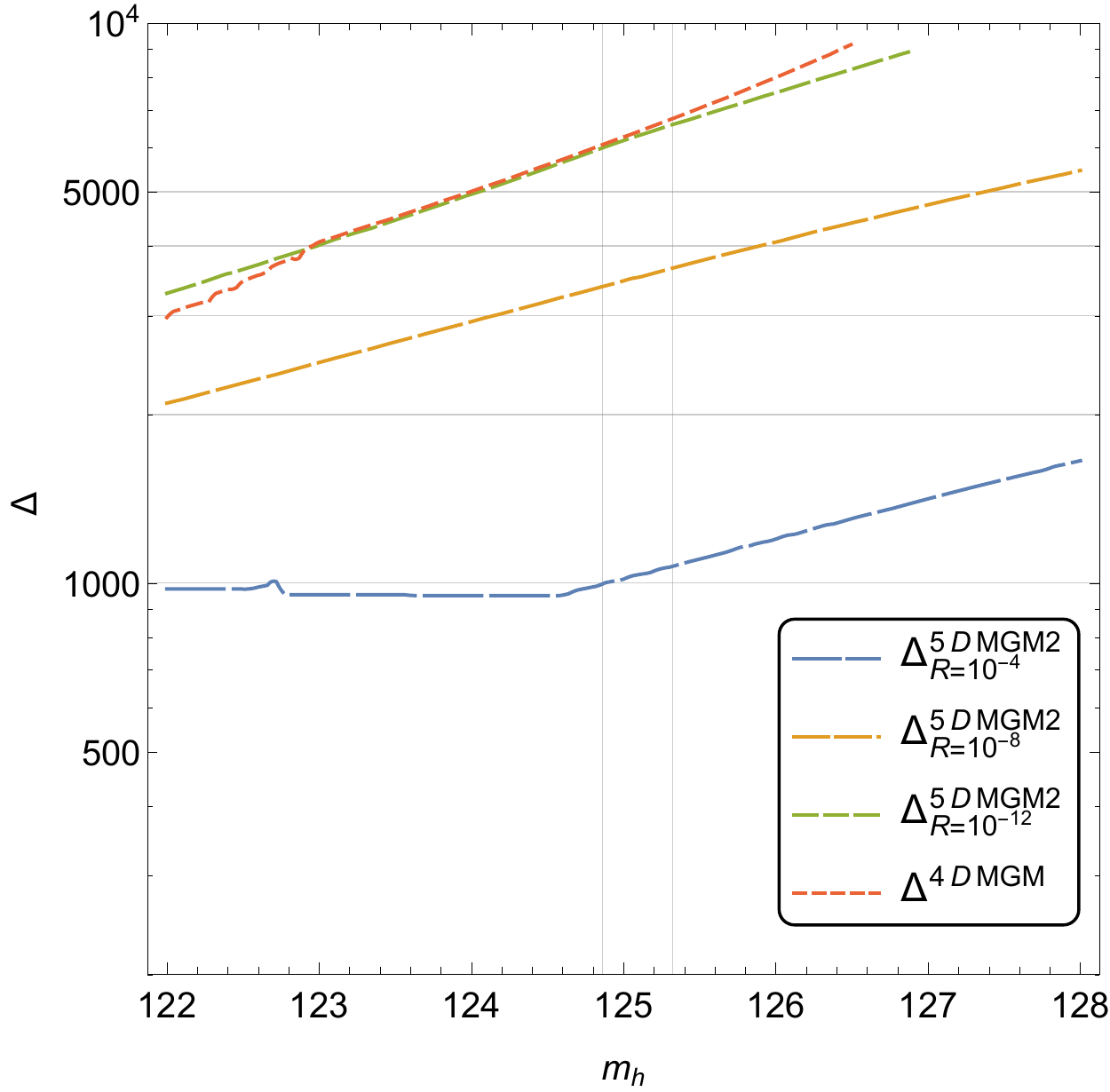}
\includegraphics[width=6.2cm]{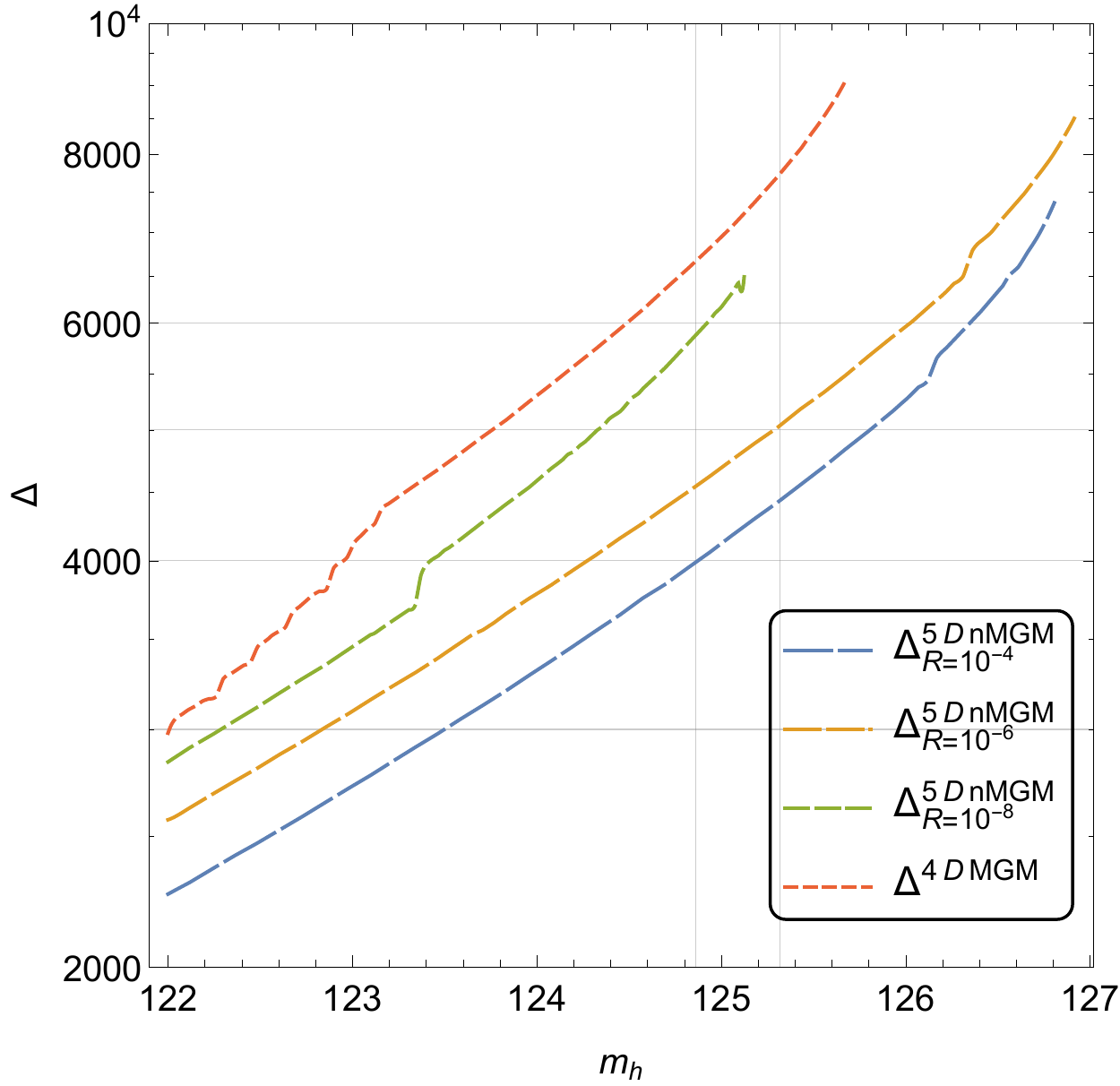}
\hspace{6.2cm}
\caption{{\it 
Fine-tuning as a function of Higgs mass for different sizes of the extra dimension for models 1 (left hand side) and 2 (right hand side) with CMSSM (top row), MGM (middle row) and nMGM (bottom row)  spectra as well as the $4D$ results.
}}  
\label{fig:FTplot}
\end{center}
\end{figure}

One might have expected a significant contribution to fine tuning from the power law contribution. However four and five dimensional theories actually have similar fine tuning as the much faster power law contribution can dominate the running for only a very small range of scales if the spectra we are comparing are similar. And so the final amount of fine-tuning for a given scenario depends mostly on the resulting spectrum rather than on the amount of power law running. This is quantified in figure \ref{fig:FTplot}, where in numerical calculations we use a standard fine-tuning measure with respect to parameter $a$ defined as
follows\footnote{The numerical procedure used to calculate fine-tuning is detailed in appendix \ref{ftappendix}.} \cite{Ellis:1985yc,Barbieri:1987fn,Chankowski:1998xv}
\begin{equation}\label{FTdef}                                                    
\Delta_a=\left| \frac{\partial \ln{M^2_Z}}{\partial \ln{a}} \right|.
\end{equation}
Fine-tuning connected with a set of independent parameters $a_i$ is then
\begin{equation}\label{Delta}                                                      
\Delta=\sum_i \Delta_{a^2_i}.  
\end{equation}
Figure~\ref{fig:FTplot} shows resulting fine-tuning as a function of Higgs mass for different  sizes of the extra dimension as well as the result one would obtain from $4D$ running. the top row shows results obtained assuming CMSSM-like soft terms (with $A_i=0$), the middle row shows gauge mediated boundary conditions and the bottom plot shows the nMGM ones. 

The results in left panel show model $1$ which gives a rather standard prediction despite power law contribution to running. However model $2$ shown on the right hand side allows us to reduce fine-tuning very significantly. The reason are the gaugino masses that decrease during 5D part of the running (as shown in Figure~\ref{fig:running2}). This protects the soft terms from the usual increase due to the heavy gluino. Since the A-terms do not grow proportionally to scalar masses we can easily achieve maximal mixing scenario for the light stops, and their direct detection bound is precisely what gives us the lower bound on fine-tuning we can see in model $2$ with $R=10^{-4}$.  

The bottom plot shows nMGM result which turns out quite similar to MGM and CMSSM model $1$ results. The reason for this is that in model $1$ the least fine tuned results are those for which $M>>1/R$. Thus the scalar masses are initially very small and have to be generated with modified running. Consequently the 3rd family part of the spectra are very similar. The correction introduced by nMGM relies only on larger subleading corrections to the Higgs mass from first two families and other Higgs sector scalars. Unfortunately fine-tuning price of these corrections is larger than their contribution to the Higgs mass and the results are slightly more fine tuned than those from standard MGM or CMSSM soft terms. 

A large qualitative difference between MGM and nMGM becomes visible for Higgs masses slightly higher than the observed one. This comes from the part of parameter space which predicts successful electroweak symmetry breaking in nMGM. As explained in the beginning of this section, the problem is a result of the exclusion appearing in nMGM for very small $1/MR$. Where we cannot break electroweak symmetry because radiative correction to the unsuppressed soft Higgs mass coming from highly suppressed stop mass is to small, and the former never runs negative. This becomes visible for higher Higgs masses because very small $1/MR$ is the part of the parameter space where we obtain highest Higgs masses.
Another very important feature of 5D models is the possibility to bring superpartner masses within the LHC reach for points predicting minimal fine-tuning. This is illustrated in Table~\ref{Table:spectra} which shows spectra corresponding to lowest obtained fine tuning for $m_h\,=\,125$\gev.
\begin{table}
\begin{center} 
\begin{tabular}{|c|c|c|c|c|c|c|c|} 
\hline 
\multicolumn{8}{  |c| }{CMSSM} \\ \hline
Model: &\multicolumn{3}{  |c| }{1}&\multicolumn{3}{  |c| }{2}& $4D$ \\ \hline 
R: & $10^{-4}$ & $10^{-6}$ &  $10^{-8}$ & $10^{-4}$ & $10^{-6}$ &   $10^{-8}$ & $4D$ \\
\hline 
$\tilde{q}_{1,2}$  &3.14 &3.45 & 3.84 &1.76&2.40&3.23&4.58  \\ 
\hline 
$\tilde{t}_1$  &2.44 &2.82 & 3.1 &0.75&1.1&2.2&3.59  \\ 
 \hline
 $\tilde{\chi}^0_1 $ & 0.85 & 1.02 &1.23&0.21&0.38&0.81& 1.26 \\ 
 \hline
 $\tilde{m}_A$  &1.82 &2.20 &2.50&2.1&2.3&2.4&2.77  \\ 
 \hline
 $\Delta / 10^{3}$  &$3.5$ &$4.0$ &$4.6$&$0.9$&$1.5$&$3$&$5.3$  \\ 
 \hline
 \hline 
\multicolumn{8}{  |c| }{MGM} \\ \hline
Model: &\multicolumn{3}{  |c| }{1}&\multicolumn{3}{  |c| }{2}& $4D$ \\ \hline 
R: & $10^{-4}$ & $10^{-6}$ &  $10^{-8}$ & $10^{-4}$ & $10^{-6}$ &   $10^{-8}$ & $4D$ \\
\hline 
$\tilde{q}_{1,2}$  &3.12 &3.45 & 3.84 &1.76&2.40&3.23&4.59  \\ 
\hline 
$\tilde{t}_1$  &2.57 &2.77 & 3.18 &0.81&1.47&2.2&3.92  \\ 
 \hline
 $\tilde{\chi}^0_1 $ & 0.80 & 0.91 &1.08&0.17&0.37&0.81& 1.32 \\ 
 \hline
 $\tilde{m}_A$  &1.82 &2.20 &2.44&1.43&1.8&1.90&2.31  \\ 
 \hline
 $\Delta / 10^{3}$  &$3.6$ &$4.0$ &$4.6$&$0.95$&$1.85$&$3.19$&$6.01$  \\ 
 \hline
 \end{tabular}
 \begin{tabular}{|c|c|c|c|c|} 
  \hline 
\multicolumn{5}{  |c| }{nMGM} \\ \hline
Model: &\multicolumn{3}{  |c| }{1}& $4D$\\
 \hline 
R: & $10^{-4}$ & $10^{-6}$ &  $10^{-8}$& $4D$\\
\hline 
$\tilde{q}_{1,2}$  &3.81 &4.41 & 5.69 & 6.41  \\
\hline 
$\tilde{t}_1$  &2.33 &2.64 & 3.15 & 3.69\\
 \hline
 $\tilde{\chi}^0_1 $ & 0.79 & 0.93 &1.01 &1.05\\
 \hline
 $\tilde{m}_A$  &1.92 &2.36 &2.92&3.31\\
 \hline
 $\Delta / 10^{3}$  &$3.7$ &$4.2$ &$5.2$&$6.0$\\
 \hline
\end{tabular} \caption{Masses of superpartners (in \tev) for spectra which minimize fine-tuning for $m_h \, =\,125$\gev 
\label{Table:spectra}}
\end{center} 
\end{table}

\section{Conclusions}\label{sec:conclude}
In this paper we explored the implementation of the five dimensional renormalisation group equations of a number of supersymmetric extensions of the MSSM, into a full C++ spectrum generator, along with self energy corrections for the Higgs mass.

Our key result is showing that modified five dimensional RGEs can result in spectra very different from the usual $4D$ case. The is because in 5D the heavy gluino does not necessarily dominate running of other soft terms during power law running, as in our model 2. Thus we can easily obtain maximal stop mixing and much less fine tuned spectra, even with standard sets of soft terms at the SUSY breaking scale.
This is also very interesting because in 5D models the least fine tuned spectra with correct Higgs mass can easily predict soft superpartner masses within LHC reach, even for standard patterns of soft terms. Interestingly, this means the most interesting parts of the parameter space can be probed during next run of the LHC, which is not usually the case in 4D models.

 We explored models where the 1st and 2nd generation are in the bulk and a model in which the 1st and 2nd generation is on the same brane as the supersymmetry breaking sector and the 3rd generation is located on an opposite brane, resulting in a spectrum of stops lighter than other squarks. Obtaining lighter stop soft terms at the SUSY breaking scale did not result in a more natural spectrum. The reason is non negligible fine-tuning price of heavier first two generations and heavier Higgs sector which give only a subleading correction to the light Higgs mass.

The final advantage is a low scale of unification of gauge couplings and a low supersymmetry breaking scale. And also much better unification of Yukawa couplings (especially in model 2) which gives hope for a very interesting five dimensional UV completion of such models.

\section*{Acknowledgements}
This work was partially supported by the Foundation for Polish Science International PhD Projects Programme co-financed by the EU European Regional Development Fund and by National Science Centre under research grants DEC-2012/04/A/ST2/00099 and DEC- 2014/13/N/ST2/02712.
ML was supported by the Polish National Science Centre under doctoral scholarship number 2015/16/T/ST2/00527.
\appendix 
\section{Numerical procedure}\label{sec:Numerical}
In this appendix we outline the implementation of the RG-solver and spectrum generator used in this paper. The numerical procedure we use is similar to the ones used in existing codes \cite{Allanach:2001kg,Porod:2003um,Djouadi:2002ze}. We work with quantities
renormalized in $\overline{DR}$ and use renormalization group equations (RGE),
to iteratively find low energy parameters for a given set of high energy soft terms.

\begin{figure}[ht]
\setlength{\unitlength}{0.25cm} 
\centering 
\begin{picture}(40,50)
\put(1,40){\framebox(35,10){
\parbox{33\unitlength}{
Calculate radiative corrections to couplings 
$g_i(M_Z)$,$h_t(M_Z)$,$h_b(M_Z)$,$h_\tau(M_Z)$
(use SM values in the first run)
}
}}
\put(15,40){\vector(0,-1){5}}\put(16,37) {$\textrm{RGE}:M_z \rightarrow M_u$}
\put(1,25){\framebox(35,10){
\parbox{33\unitlength}{
Include soft breaking terms given at high scale $M_u$. Run from $M_u \rightarrow 1/R$ with 5D RGEs and $1/R\rightarrow M_{EWSB}$ with four dimensional RGEs.
}
}}
\put(15,25){\vector(0,-1){5}}\put(16,22) {$\textrm{RGE}:M_u \rightarrow
M_{EWSB}$}
\put(1,10){\framebox(35,10){
\parbox{33\unitlength}{
Iteratively calculate $\mu$,$B_\mu$ and the mass spectrum
(in the first run find estimates for $M_{EWSB}$
,$\mu$ and $B_\mu$)
}
}}
\put(15,10){\vector(0,-1){5}}\put(16,7) {if $\mu$ converged}
\put(1,0){\framebox(35,5){
\parbox{33\unitlength}{
Calculate physical masses
}
}}
\put(36,15){\line(1,0){2}}\put(36,45){\line(1,0){2}}
\put(38,15){\vector(0,1){10}}\put(38,15){\line(0,1){30}}
\put(39,22) {
\begin{sideways}
$\textrm{RGE}:M_{EWSB} \rightarrow M_Z$
\end{sideways}
}
\end{picture}
\parbox{40\unitlength}{
\caption{
Schematic of the numerical algorithm. Subsequent steps are described in the appendix.
} 
}
\label{algorytm}
\end{figure}
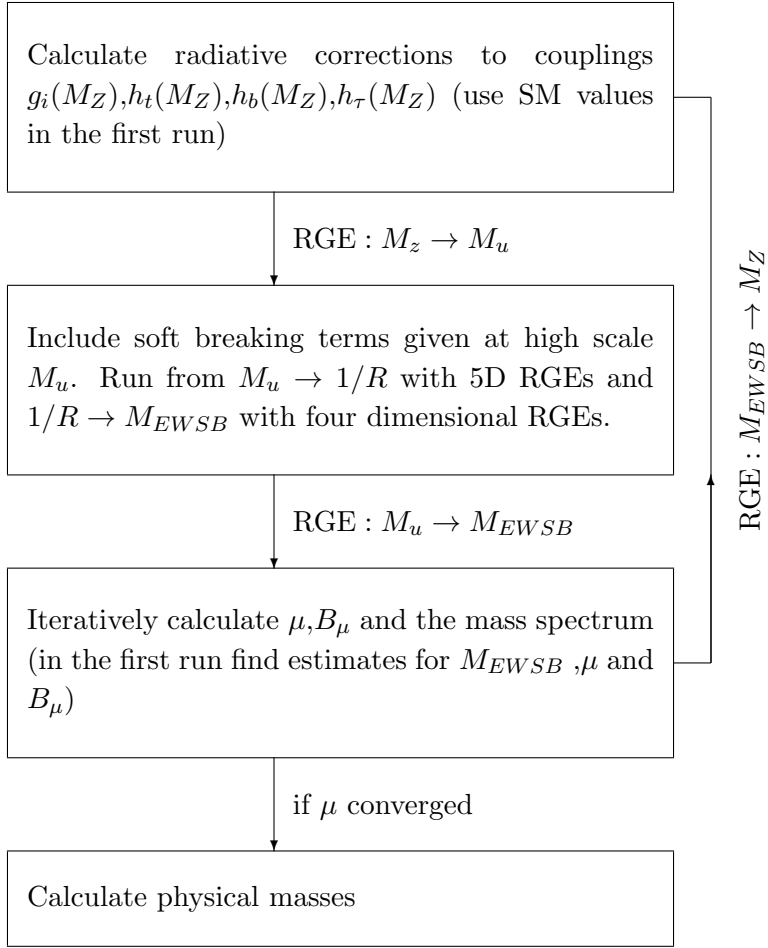
\subsection{$M_Z$ Scale}
At the scale $M_Z$ we include radiative corrections to couplings. We set Yukawa couplings using the tree-level relations
\begin{equation}
y_t = \frac{m_t \sqrt{2}}{v \sin{\beta}} \quad , \quad
y_b = \frac{m_b \sqrt{2}}{v \cos{\beta}} \quad , \quad
y_{\tau} = \frac{m_{\tau} \sqrt{2}}{v \cos{\beta}},
\end{equation}
where $m_t , m_b , m_{\tau}$ are fermion masses and $v$ is the Higgs vev. At  the first iteration we use physical masses and SM Higgs
vev $v\approx 246,22$\gev. During subsequent iterations above quantities are renormalized in
$\overline{DR}$ scheme and one-loop SUSY corrections are included.
To calculate the top mass we use 2-loop QCD corrections \cite{Avdeev:1997sz}
and 1-loop corrections from super-partners from the appendix of
\cite{Pierce:1996zz}. While calculating the bottom mass we follow \textit{Les Houches Accord} 
\cite{Skands:2003cj}, starting from running mass in $\overline{MS}$ scheme in SM
${m_b}^{\overline{MS}}_{SM}$. Next applying the procedure described in \cite{Baer:2002ek}
we find  $\overline{DR}$ mass at $M_Z$, from which we get MSSM value by
including corrections described in appendix D of \cite{Pierce:1996zz}.
While calculating the tau mass we include only leading corrections approximated in \cite{Pierce:1996zz}.
We calculate the Higgs vev in the MSSM using
\begin{equation}
v^2=4 \frac{M^2_Z+\Re{\Pi^T_{ZZ}(M_Z)}}{g^2_2+3 g^2_1 /5},
\end{equation}
where we include $Z$ self interactions described in appendix D of \cite{Pierce:1996zz}.
To calculate $g_1$ , $g_2$, $g_3$ in $\overline{DR}$ in the MSSM we use the procedure described in appendix C of \cite{Pierce:1996zz}.
\subsection{RGE and $M_u$ scale}
After calculating coupling constants at the scale $M_Z$ we numerically solve RGEs 
\cite{Martin:1997ns},\cite{Yamada:2001ck}, to find their values at the scale $M_u$, at which we include the soft breaking terms. Then we solve RGEs again to find soft terms, coupling constants, $\tan{\beta}$ and Higgs vev
$v$ at the scale
$M_{EWSB}=\sqrt{m_{\tilde{t}_1}(M_{EWSB})m_{\tilde{t}_2}(M_{EWSB}})$. 
At first iteration we take  $\mu=\textrm{sgn}(\mu) \  \textrm{GeV}$
and $B_\mu=0$ and run to the scale at which the above equation is fulfilled.
\subsection{Electroweak symmetry breaking}
In order to obtain correct electroweak symmetry breaking we use
minimization conditions for the scalar potential to find new values of $\mu$ and $B_\mu$.
We include radiative corrections in these equations by the substitution
\begin{equation}
m_{H_u} \rightarrow m_{H_u}+\frac{t_u}{v_u} \quad , \quad
m_{H_d} \rightarrow m_{H_d}+\frac{t_d}{v_d}.
\end{equation}
We include full one-loop corrections to $t_u$ and $t_d$
presented in  appendix E of \cite{Pierce:1996zz} and leading two-loop corrections
\cite{Dedes:2002dy,Dedes:2003km,Brignole:2002bz,Brignole:2001jy,Degrassi:2001yf}. Since these corrections depend on sparticle masses 
which in turn depend on the $\mu$ parameter that we aim to calculate, an iterative
calculation is performed to obtain new values of $\mu$ and $B_\mu$.

If the new values differ significantly from the ones obtained in previous repetition
of the whole algorithm described above, we run back to the 
$M_Z$ scale and repeat the whole calculation once again.
If however the values of $\mu$ and $B_\mu$ converged, we can move on to teh calculation of physical masses. 
\subsection{Calculation of physical masses}
To calculate physical masses we use only leading corrections described in \cite{Pierce:1996zz} everywhere but the Higgs sector.
In the Higgs masses calculation we use full one-loop corrections from
\cite{Pierce:1996zz} and leading two-loop corrections described in \cite{Dedes:2002dy,Dedes:2003km,Brignole:2002bz,Brignole:2001jy,Degrassi:2001yf}. 
 \subsection{Fine-tuning}\label{ftappendix}
After the calculation of the spectrum is finished, one has a whole set of parameters and couplings that predict correct electroweak symmetry breaking. In order to calculate fine-tuning we solve the RGEs from $M_u$ scale down to $M_{EWSB}$ with one of the fundamental parameters $a_i$ changed slightly at the high scale $M_u$ . Than at the scale $M_{EWSB}$ we recalculate the spectrum and  use minimization conditions to calculate a new  value of $\tan \beta$ and  to obtain our new prediction for $m^2_Z$, which means that  we calculate numerically  the derivative in the definition of fine-tuning \eqref{FTdef}. We repeat that procedure for all parameters $a_i$ and obtain our final result as a maximum of results obtained  for each of those parameters (as in \eqref{Delta}). 

\bibliographystyle{JHEP}
\bibliography{fiveD}

\end{document}